\documentclass[authoryear]{elsarticle}

\usepackage{graphicx}
\usepackage{amssymb}
\usepackage{gensymb}

\usepackage{lineno}

\usepackage{setspace}

\usepackage{caption}
\usepackage{siunitx}

\newcommand\textmicrometer{~{\textmu}m}
\newcommand\micrometer{%
\ifmmode\textnormal{\textmicrometer}%
\else\textmicrometer%
\fi}

\usepackage{geometry}
 \usepackage{hyperref}
 
\usepackage{eucal}
 
\journal{Remote Sensing of Environment}

\begin{document}

\begin{frontmatter}


\title{Estimating heterogeneous wildfire effects using synthetic controls and satellite remote sensing}

\doublespacing


\author[a,b,*]{Feliu Serra-Burriel}
\author[b,d]{Pedro Delicado}
\author[a,c]{Andrew T. Prata}
\author[a]{Fernando Cucchietti}
\address[a]{Barcelona Supercomputing Center, Barcelona, Spain}
\address[b]{Department of Statistics and Operations Research, Universitat Polit\`ecnica de Catalunya, Barcelona, Spain}
\address[c]{Sub-department of Atmospheric, Oceanic and Planetary Physics, University of Oxford, Oxford, UK}
\address[d]{Institut de Matem\`atiques de la UPC - BarcelonaTech (IMTech), Barcelona, Spain}
\address[*]{Corresponding author: feliu.serra@bsc.es}


\begin{abstract}
Wildfires have become one of the biggest natural hazards for environments worldwide. The effects of wildfires are heterogeneous, meaning that the magnitude of their effects depends on many factors such as geographical region, climate and land cover/vegetation type. Yet, which areas are more affected by these events remains unclear. Here we present a novel application of the Generalised Synthetic Control (GSC) method that enables quantification and prediction of vegetation changes due to wildfires through a time-series analysis of in situ and satellite remote sensing data. We apply this method to medium to large wildfires ($>$ 1000 acres) in California throughout a time-span of two decades (1996--2016). The method's ability for estimating counterfactual vegetation characteristics for burned regions is explored in order to quantify abrupt system changes. We find that the GSC method is better at predicting vegetation changes than the more traditional approach of using nearby regions to assess wildfire impacts. We evaluate the GSC method by comparing its predictions of spectral vegetation indices to observations during pre-wildfire periods and find improvements in correlation coefficient from $R^2 = 0.66$ to $R^2 = 0.93$ in Normalised Difference Vegetation Index (NDVI), from $R^2 = 0.48$ to $R^2 = 0.81$ for Normalised Burn Ratio (NBR), and from $R^2 = 0.49$ to $R^2 = 0.85$ for Normalised Difference Moisture Index (NDMI). Results show greater changes in NDVI, NBR, and NDMI post-fire on regions classified as having a lower Burning Index. We find that on average, wildfires cause a 25\% initial decrease in the vegetation index (NDVI) and a larger than 80\% drop in wetness indices (NBR and NDMI) after they occur. The GSC method also reveals that wildfire effects on vegetation can last for more than a decade post-wildfire, and in some cases never return to their previous vegetation cycles within our study period. We also find that the dynamical effects vary across regions and have an impact on seasonal cycles of vegetation in later years. Lastly, we discuss the usefulness of using GSC in remote sensing analyses.
\end{abstract}


\end{frontmatter}

\doublespacing
\section{Introduction}
Wildfires pose a significant natural hazard to society \citep{paton2015}. Moreover, the increase in the frequency and intensity of extreme weather and climate events lead to an increase in societal vulnerability to wildfires \citep{Easterling2000, Moritz2014, Schoennagel2017}.
Climate change is expected to increase the amount of areas at risk of large wildfires \citep{Westerling2011}. Given the rise in temperatures, and a future drier climate, as climate projections show \citep{Westerling2006,Spracklen2009, Bryant2014, Schoennagel2017, Angelo2017}, together with changes in the timing of seasons \citep{Westerling2016}, the situation is expected to worsen \citep{Littell2018}. 
Wildfires are discrete events with a strong seasonality \citep{NationalFireDataCenterU.S.2005}, but changes in seasons have altered the periodicity and effects of wildfires \citep{Westerling2006, Jolly2015, Westerling2016}. In order to mitigate the hazards posed by wildfires it is important to quantify post-fire vegetation recovery and loss, allowing for effective short- and long-term land management \citep{Chu2013}. Predicting the timeline and capacity for fire-prone regions to return to their pre-fire or an alternative state is also desirable, as post-fire vegetation recovery can lead to a significant carbon sink that can offset carbon losses caused by wildfire events \citep{Hicke2003, Meng2018}.

Climate is generally considered to vary gradually over multi-decadal timescales. However, anthropogenic climate change has been demonstrated to be associated with identified climatic presses and pulses causing complex and catastrophic responses~\citep{Harris2018}. In terms of wildfires, anthropogenic climate change has been attributed to increasing their size and frequency \citep{Abatzoglou2016, Williams2019}.
California (USA) is a region of particular interest due to the fire-proneness of the state and value of its national and state parks. Wildfire impacts vary across the state due to the north-south climate gradient \citep{minnich2018california, Goulden2012} and variation in altitude \citep{Casady2010} and land cover types. In recent years, California has endured some of the largest wildfires in history \citep{calkin2020california} and so it is important to develop methods that quantify and predict vegetation recovery patterns following large wildfire events in this region. State-wide time-series analyses of burned areas in California are now possible due to the significant amount of freely available data provided by the Monitoring Trends in Burn Severity (MTBS) \citep{Eidenshink2007} program, conducted by the U.S. Geological Survey Center for Earth Resources Observation and Science (EROS) and the USDA Forest Service Geospatial Technology and Applications Center (GTAC).
The likely increase of large wildfire frequency and severity \citep{Bryant2014} necessitates new methods using dynamical models to estimate their effects on vegetation recovery patterns.  Understanding which vegetation type is most affected by these events is essential. In addition to in situ data, time-series analysis of satellite remote sensing data can reveal more subtle changes in ecosystem health and conditions \citep{Li2020}.

Given the complexity of disentangling factors known to influence vegetation and forest composition \citep{Zhang2018ndvi} such as climate-induced changes in meteorology (e.g.  solar radiation, temperature, and precipitation) and land surface characteristics, long-term, high resolution geospatial datasets are required to estimate and predict the cost of wildfires on vegetation \citep{DALE2001, Meng2018, Sturrock2011, Seidl2017}. Satellite remote sensing has become an invaluable tool for monitoring and assessing wildfire activity \citep{geist2005our, chuvieco2012remote, Meng2018}. In addition, the archive of satellite data suitable for land cover and wildfire assessment spans more than 30 years. 

To estimate the effects of wildfires, previous studies have used satellite observations to make comparisons of trends between similar burned and unburned regions \citep{Goetz2006, Alcaraz-Segura2010, Bolton2015}. 
\citet{Bright2019} found that wildfires in temperate forest ecosystems have diverse effects, with some regions taking less than 5 years to more than 13 years to recover pre-wildfire vegetation indices values. 
However, most of previous analyses did not consider the initial vegetation conditions, nor the differences in forest ecosystem, vegetation type and climate. 
As the natural environment varies over time, there can be differences between control regions and burned regions in terms of natural gradients such as diversity, fertility and soil moisture \citep{Ibanez2019} that need to be accounted for. 
Previous studies have used differences between pre- and post-wildfires through satellite imagery to compute a post-disturbance regrowth, in both, absolute (growth trend) and relative (recovery indicator) terms \citep{Kennedy2012}.
Other studies used control regions as counterfactual vegetation to estimate the decrease in gross primary production (GPP) of terrestrial vegetation after a wildfire \citep{Steiner2020}. 
Spectral similarities between affected and unaffected pixels for change detection of ecosystem dynamics on time series have also been explored \citep{Lhermitte2010}.
The most detailed methodologies require the combination of remotely sensed and field ground data \citep{Chu2013}.
However, most of these studies focus on small groups of wildfires that have large impacts on vegetation, with small pre- and post-fire follow-up dynamics. 
In addition, to our knowledge, no methodology accounts for time-varying confounding factors and non-stationary, vegetation index time series data to estimate the dynamical heterogeneous effects of wildfires. 
This is an important consideration as decadal increases or decreases in NDVI have been shown to impact analysis of historical wildfire case studies \citep{Hicke2003}. 
Moreover, these studies are limited by the heterogeneities of the unburned areas used, as they might differ from the heterogeneities in burned regions.  

Here we propose a novel approach for estimating the impact of wildfires using the generalised synthetic control (GSC) method \citep{Xu2017}. The GSC method was originally developed for estimating the effects of government policies and can be used to measure causal effects from interventions.
In this paper we apply the GSC method to long-term ($\sim$30 years) time-series of satellite-derived vegetation health indices (i.e. NDVI, NDMI, NBR). In particular, we are most interested in the NDVI, as it outperforms other spectral indices' accuracy in areas with heterogeneous vegetation and it is the most robust vegetation index for assessing vegetation recovery \citep{Veraverbeke2012}. 
Specifically, we consider the surroundings of burned regions to reconstruct the characteristics of vegetation that would have been observed in the absence of a wildfire (referred to hereafter as control areas or control pixels \citep{Veraverbeke2010}). These regions, as well as the burned areas, are represented as polygons, and the information is aggregated from pixel-level data and averaged to obtain spatial and temporal spectral indices for each of the areas of interest (AOIs). 
The GSC method also allows us to take into account weather and climate data. As a consequence, we are able to detect decreases in post-fire vegetation seasonal-cycles and estimate vegetation recovery times. 

\section{Materials and Methods}\label{sec:matmethods}

\subsection{Study Area}

The study region considered for the present analysis is the state of California, USA (Fig.~\ref{fig:MAP_cali_aois}), which is within approximately 32--44{\textdegree}N and 112--126{\textdegree}W. The vegetation in this region is predominately classified as shrublands, grasslands, and evergreen forests \citep{Jin2019}. 
California has a largely Mediterranean climate but encompasses regions that vary from hot desert to alpine tundra. The wildfire season generally occurs between May and October when weather conditions in the western United States are hot and dry \citep{Westerling2003}. According to the GridMET interpolated surface meteorological dataset \citep{Abatzoglou2013}, between 1990 and 2018, in the summer months (June-August), the statewide climatological average maximum and minimum temperatures were 31.04 {\textdegree}C and 14.7 {\textdegree}C, respectively. Most of the rainfall occurs in winter (climatological average precipitation of 3.59 mm) while the summer is much drier (climatological average precipitation 0.21 mm), resulting in high wildfire ignition risk in many areas during the summer months \citep{Bryant2014}.

\subsubsection{Burned Areas}

The MTBS dataset contains 1631 wildfires that burned areas larger than 1000 acres in California between 1984 and 2016. The reason behind this cutoff is that the MTBS dataset only contains wildfires with perimeters larger than 1000 acres ($\sim 405 hectares)$. Out of these 1631 burned areas, 544 only burned once, meaning that the perimeters from these fires were non-overlapping. However, as we explain in the next sections, we require $\sim$10 years of data from pre-wildfire periods to compute synthetic controls and so we only consider the effects of wildfires after the year 1995. Hence, from 1996--2016, there were 342 burned areas that only burned once (Fig. \ref{fig:MAP_cali_aois}). From these 342 fires, 22 were prescribed fires, 7 were catalogued as unknown source, and 6 were 'wildland fire use' according to the MTBS \citep{Eidenshink2007}, that is, natural fires allowed to burn when outlined in a fire management plan and communities are not at risk. The rest of them (307) were catalogued as wildfires and are used in our analysis together with their control areas to define our AOIs (indicated as red and green polygons in Fig.~\ref{fig:MAP_cali_aois}). The wildfires span a total area of approximately $5 \times 10^5$ hectares. The temporal pattern of wildfire occurrence follows an approximately constant number of wildfires over years, most of these during summer period when fuels are ready to spark, except for a large spike on 2008 with some of the largest wildfires in California’s history. Most of these wildfires occurred in the center and northern non-coastal part of California and in places where the most predominant land cover was shrub/scrub, grassland herbaceous, or evergreen forest.

\subsubsection{Control Areas}

\begin{figure}[h]
\includegraphics[scale=.35]{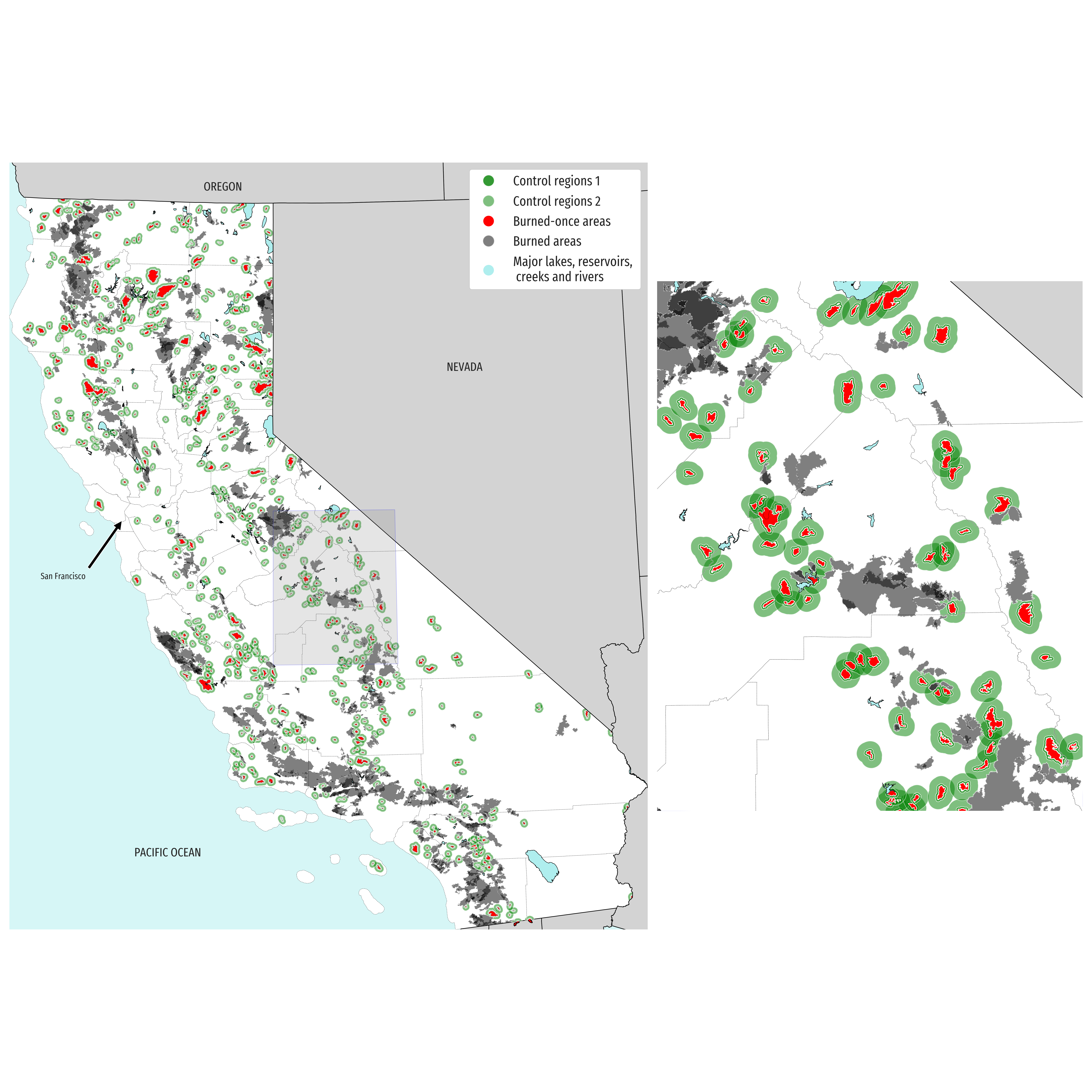}
\caption{\textbf{Map of California and the areas of interest. }
The map shows the regions used in this study. Shown in red are the areas that burned only once during the studied period. In shades of green are shown the two non-overlapping control regions. Gray regions represent areas that suffered from overlapping wildfires, and thus were not included on the estimation of wildfire disturbance effects, and darker shades of gray indicate the parts of the areas of the burned perimeters that overlay with more than one wildfire perimeter, specifying the exact areas that suffered multiple burns.}
    \label{fig:MAP_cali_aois}
\end{figure}

We generated control areas (or ``buffer zones'') around the burned areas (e.g. Controls 1 are the polygons enclosing burned areas from 100 m to 1 km away from the perimeter of the burned areas and Controls 2 are from 1 km to 5 km away, as shown in Fig.~\ref{fig:MAP_cali_aois}), similar to \citet{Goetz2006}, used to estimate counterfactual vegetation indices. In addition, we removed any pixels from control regions that intersected neighbouring burned areas overlapping control region boundaries from our analysis, to avoid biasing estimates of counterfactual vegetation.

As we will demonstrate in the next sections, the methodology proposed here shows that the combination of aggregated pixels from control areas together with burned regions is suitable for the estimation of wildfire effects, as the estimation of counterfactual vegetation uses information from all control regions on post-wildfire periods, as well as information from both control and burned regions on pre-wildfire periods. The use of different control regions ensures that the effects are unaltered by potential spillover wildfire effects on the control regions.

\subsection{Landsat surface reflectance spectral indices and climate data}

The spatio-temporal study of vegetation dynamics over large time-spans and areas on complex environments and ecosystems is mostly accomplished with remote sensing, with satellites consistently capturing multi-spectral data over long time periods. 
Spectral indices are commonly used for monitoring and exploring vegetation dynamics around the globe~\citep{Hislop2018,Kennedy2010}. 
The most common spectral index used in the literature is the Normalized Difference Vegetation Index (NDVI)~\citep{rouse1974monitoringNDVI}, which is calculated as the difference between the near-infrared (centred near 
\SI{0.87}{\micro\metre}) and red (centred near \SI{0.66}{\micro\metre}) reflectance over their sum (Table~\ref{table:LSRDSI}), and correlates with biomass and ecological outcomes, and it has been widely used to monitor vegetation and detect forest disturbances. 
Another spectral index that is widely used in the remote sensing literature for assessing post fire vegetation dynamics is the Normalized Burn Ratio (NBR)~\citep{key1999normalizedBR}, which is analogous to the NDVI but exploits the near-infrared and second short-wave infrared (SWIR 2; centred near \SI{2.2}{\micro\metre}) channels. This index has been shown to correlate highly with ground-based measurements and to be a reliable tool for post-fire vegetation dynamic assessment. Lastly, the Normalized Difference Moisture Index (NDMI)~\citep{wilson2002detectionNDMI} is the same as the NBR but uses the first Landsat short-wave infrared channel (SWIR 1; centred near \SI{1.61}{\micro\metre}). It is similar to the NBR in the sense that it correlates with field-based measurements of vegetation, however, this spectral index is used for other non-fire related abrupt changes. 

Using Landsat Surface Reflectance Derived Spectral Indices (LSR-DSI) time series data \citep{Bolton2015}, extracted with the Google Earth Engine platform (GEE) \citep{gorelick2017google}, we construct NDVI, NDMI and NBR indices for three Landsat satellites (LT5, LT7 and LO8) \citep{Cohen2017, Roy2016} masking clouds, shadows and snow pixels with FMASK \citep{Zhu2012}, and removing pixels from water bodies such as lakes, reservoirs, rivers and creeks. The Landsat satellites provide a consistent source of 30~m per pixel resolution, with a frequency of 16 days. The data was collected using GEE from the beginning of 1990 until the end of 2018. Table~\ref{table:LSRDSI} summarises the indices that have been used for this study, together with citations to studies used as reference for this work. 
\begin{table}[h]
\centering
\begin{tabular}{r|c|c}
\textbf{Greenness Indices} & \textbf{Formulas} &  \textbf{P-B. TS Studies} \\ \hline 
\hline

\begin{tabular}{r} Normalized Difference\\[-.3em] Vegetation Index (NDVI)\end{tabular} & $ NDVI = \frac{NIR-RED}{NIR+RED}$&      \citep{rouse1974monitoringNDVI, pettorelli2005usingNDVI}  \\
\hline

\textbf{Wetness Indices}   & \textbf{ }   &  \textbf{ } \\ \hline
\hline
\begin{tabular}{r} Normalized Burn\\[-.3em] Ratio (NBR)\end{tabular}    & $NBR = \frac{NIR-SWIR2}{NIR+SWIR2}$         &                     \citep{key1999normalizedBR}, \citep{Eidenshink2007}             \\
\begin{tabular}{r} Normalized Difference\\[-.3em] Moisture Index (NDMI)\end{tabular}             & $NDMI = \frac{NIR-SWIR1}{NIR+SWIR1}$       &          \citep{wilson2002detectionNDMI}    

\end{tabular}
\caption{Landsat Surface Reflectance - Derived Spectral Indices used in this work and a selection of pixel-based time-series studies using these. The names provided in the formulas, Near Infra-Red (NIR), Short Wave Infra-Red (SWIR1 and SWIR2) and RED, correspond to the bands that capture different wavelengths from the spectrum.}
\label{table:LSRDSI}
\end{table}

This enables us to obtain a multivariate spatial time series over the regions of interest in the state of California, USA--namely, the regions that burned only once during the observed two-decade period from 1996--2016 with their respective control areas. We have narrowed down the areas of interest to single-burned regions because we want to estimate causal effects, and the methodology only allows for binary treatment (in the context of this paper, the term \emph{treatment} refers to a wildfire event, and \emph{units} are defined as any particular AOI). The outcome is observed for a longer period (1990-2019), however, these methods require a reasonably long time period before the wildfire occurs, such that the matching of treated and control regions is sufficient. Different pre-fire periods were used considering that at least 3 years are required to capture periodicities and trends, as it is the most elementary cycle of the meteorological element \citep{Han2011}. 

Ecological land-surface models generally require climatological variables as inputs, such as air temperature, humidity, precipitation and incident solar radiation \citep{Running1987}. For this purpose, climate data was obtained from the interpolated surface meteorological data GridMET \citep{Abatzoglou2013} through GEE as well, to account for precipitation amount (daily total in milimeters), maximum temperature (daily maximum in Kelvins), 1000 hour accumulated dead fuel moisture (as a percentage) and solar radiation or surface downward shortwave radiation (in $	W/m^2$) over the desired time period for each area of interest. We also use the burning index (BI) from GridMET, which is the National Fire Danger Rating System (NFDRS) fire danger index, \citet{Wildfire2002}, as a proxy for fire weather hazard, in order to stratify observations to study how vegetation recovery patterns change with varying degrees of fire weather hazard.
Due to human causes, high fire weather hazard is not necessarily linked to actual fire hazard. However, even though this BI is not a perfect estimation of the probability of wildfire occurrence, we consider this as a reliable approximation for our purposes. Table \ref{table:GridMet_table} summarises the variables obtained from GridMET and the units they are measuring.

\begin{table}[h]
    \centering
    \begin{tabular}{r|c}
        \textbf{Climate variables} & \textbf{Units} \\ 
        \hline \hline 
         Daily precipitation & $mm$ \\ \hline
         Surface downward shortwave radiation & $W/m^2$ \\ \hline
         Maximum temperature & $K$ \\ \hline
         Burning index (NFDRS) & No units \\ \hline

    \end{tabular}
    \caption{Climate variables used in this study from the GridMET dataset \citep{Abatzoglou2013}, with a spatial resolution of $4km \times 4km$ obtained in March 2021.}
    \label{table:GridMet_table}
\end{table}

\subsection{Statistical analysis}
To estimate the heterogeneous effects of wildfires within our study region and time-period, we propose a GSC methodology that uses the Nuclear Norm Matrix Completion Method (NNMCM) \citep{athey2018matrix}. Analogous to synthetic control methods, the GSC with NNMCM creates synthetic controls for the estimation of vegetation over a region in the absence of a wildfire. This method creates synthetic observations that fit the vegetation dynamics on pre-wildfire periods, and extrapolates it to the post-fire periods, using information from all control regions in post-wildfire time periods. As a result, this allows us to estimate the differences between the observed burned vegetation and the estimated counterfactual vegetation. This methodology also accounts for time-varying factors (such as decadal variations in spectral index) and allows us to approximate the effect of each wildfire separately.

\subsubsection{Synthetic Controls}

The GSC method \citep{Xu2017} is defined based on the relation between the studied outcome (e.g. NDVI, NBR or NDMI), the observed covariates (e.g. maximum temperature, daily accumulated precipitation, solar radiation and accumulated dead fuel moisture), and treatments (wildfires) in a functional form. Following the notation of \citet{Xu2017}, let us define $Y_{it}$ as observation $i$ at time $t$ for outcome $Y$ so that
\begin{equation}
    Y_{it} = \delta_{it}D_{it} + {x_{it}}'\beta + {\lambda_{i}}'f_t + \epsilon_{it},
\end{equation}
where $Y_{it}$ is the observed spectral index of interest, $D_{it}$ is a binary variable indicating whether observation $i$ was burned before time period $t$, $\delta_{it}$ is the estimated effect for observations $i$ at time period $t$, $x_{it}'$ is a transposed matrix of observed covariates, $\beta$ is a vector of unknown parameters, $f_t$ is a vector of unobserved common factors, $\lambda_i'$ is a transposed vector of unknown factor loadings and $\epsilon_{it}$ is a matrix containing the error terms.

In order to formalize the notion of causality \citep{RubinD.B1974, holland1986statistics, Rosenbaum2006}, and following the above notation of outcomes on $Y_{it}$, we define two sets, 
$\CMcal{T}$ and $\CMcal{C}$, as the sets in \emph{treatment} and \emph{control} groups respectively.
Then, the total number of observations is $N=N_{cr}+N_{tr}$, where $N_{tr}$ is the number of treated observations (areas that burned once), and $N_{cr}$ is the number of untreated observations (control areas that never burned). 
The time variable $t$ is composed of two parts. 
The first component, $t \in \{1, 2, \cdots, T_{0}^{i}\}$, contains the pre-wildfire periods, where $T_{0}^i$ is the time of the wildfire occurrence for observation $i$.
The second component, $t \in \{T_{0+1}^i, \cdots, T_{T}^{i}\}$, contains the number of periods observed after the wildfire occurred for each AOI $i$. 
We define $\tau$ as the minimum number of pre-treatment periods to consider a wildfire, and we collect all observations $i$ that have $t \in \{1^{i}, ..., T_{0}^{i}\}$, where $T_0^i > \tau$. 

We now need to introduce the concept of potential outcome. We define $Y_{it}(1)$ as the outcome observed when units are treated (for $t > T_{0}^{i}$), and $Y_{it}(0)$ as the potential outcome, which cannot be observed by definition on treated units. That is, what would have happened in the absence of treatment, or in our case, a wildfire. Next, we shift the observation times for all the units that suffered a wildfire, so that all $T_{0}^{i}$ occur at the same time, $T_{0}$. In this way, we formulate the average treatment effect on treated as \citep{Xu2017}:

\begin{equation}
    ATT_{t, t> T_{0}} = \frac{1}{N_{tr}} \sum_{i \in \CMcal{T} } [Y_{it}(1) - Y_{it}(0)] = \frac{1}{N_{tr}} \sum_{i \in \CMcal{T}} \delta_{it} 
\end{equation}

Hence, our estimate of the treatment effect on treated unit $i$ at time $t$ is given by the difference between the actual observed outcome and its estimated counterfactual $\hat{Y}_{it}$:

\begin{equation}
    \hat{\delta}_{it} = Y_{it}(1) - \hat{Y}_{it}(0).
\end{equation}

Several assumptions are important for the notion of causality. The first one is that we need to be able to define the relation between vegetation health indices and the observed covariates and treatments in a functional form, as shown above.
Second, we need to ensure parallel trends on pre-treatment periods between treated and untreated units. Third, treatment is considered to be binary. Fourth, we need to assume regularity conditions.
Lastly, there is no spatial dependence. That is, we are assuming that treatments are assigned randomly. In order to relax this assumption, we have conditioned on the probability of wildfire to obtain conditional average treatment effects on treated (CATTs). Further discussion on these assumptions can be found in discussion section. The conditional ATT (CATT) is defined as 

\begin{equation}
    CATT = \mathbb{E}[Y_{it}(1) - Y_{it}(0)|t>T_0, X]    
    \label{formula:CATT}
\end{equation}

where conditional on the value of the observed covariate $X$, we can infer the expected value of the difference between potential outcomes. 

Figure ~\ref{fig:MAP_cali_aois} shows that the occurrence of wildfires in California is not spatially homogeneous \citep{Li2020}, and thus, the probability of different areas burning is unequal. To evaluate CATT in the present study, we propose to condition on the BI obtained from the GridMET interpolated data over the AOIs. That is, we want to estimate the average effect of wildfires on burned regions conditional on the BI. We segregate groups of observations based on observable characteristics, or clusters of similar probabilities of large wildfire incidence. Wildfires are grouped into 4 different categories using the quartiles of the BI on pre-wildfire periods. The goal is to find groups of wildfires, such that within those areas, before a wildfire occurs, the probability of a wildfire occurring in any given year is close to being random.

\subsubsection{Matrix Completion}

The matrix completion method \citep{athey2018matrix}, develops from the matrix completion literature \citep{Candes2009, candes2010matrix} a method that allows for the exploitation of both stable patterns over time and stable patterns across units.
By means of using the nuclear norm matrix completion estimator, we can obtain estimates of the missing values in the desired outcomes, as well as the estimates of the counterfactual outcomes for each of the treated units, $\hat{Y}_{it}(0)$. 

For simplicity, ignoring that we have covariates, what we have is a matrix of observed outcomes (in our case we have the above mentioned spectral indices). Thus, we can model a $N \times T$ matrix of outcomes $Y$, where $N$ is the number of observations and $T$ is the number of time periods, as:
\[
Y = L^*+\epsilon
\] 
where
\[
\mathbb{E}[\epsilon | L^*]=0
\]
with $L^* \in \mathbb{R}^{N \times T}$ and $\epsilon_{it}$ is understood as measurement error. 
Using this technique we can estimate not only the missing values previous to the treatment periods, but also the potential outcomes after the treatment is applied.
The use of non-treated observations similar to pre-treatment periods of treated units, as well as information from control regions on post-wildfire periods, allows for accurate estimates of $\hat{Y}_{it}(0)$, calibrating accurately the model on pre- and post-wildfire periods. 

Lastly, we define $\Omega$ as a set of pairs of indices $(i, t), i \in \{1, \cdots, N\}, t \in \{1, \cdots, T\}$ of the observed outcomes. One way to formulate the objective function to estimate is as follows:
\[
\hat{L} = \arg \min_L \{\frac{1}{|\Omega|} \sum_{(i,t) \in \Omega} (Y_{it}-L_{it})^{2} + \lambda_L||L||_*\}
\]
where $\lambda_L$ is chosen by cross-validation.

To estimate the uncertainty of the estimated $ATT_t$, a nonparametric bootstrap technique is used. In order to obtain a bootstrapped sample with the same number of treated and control observations as the original one, a random sample with replacement is applied to all treated and control observations separately. Then, the Matrix completion synthetic control method is applied. This procedure is repeated $B$ times.
The standard error of the estimated $ATT_t$ is computed as the sampling standard deviation of the $B$ bootstrapped $ATT_t$, and it is used to define confidence intervals based on a normal distribution approximation.

\section{Results}\label{sec:results}

\subsection{Average effect of wildfires}

\begin{figure}[h]
    \includegraphics[scale=.45]{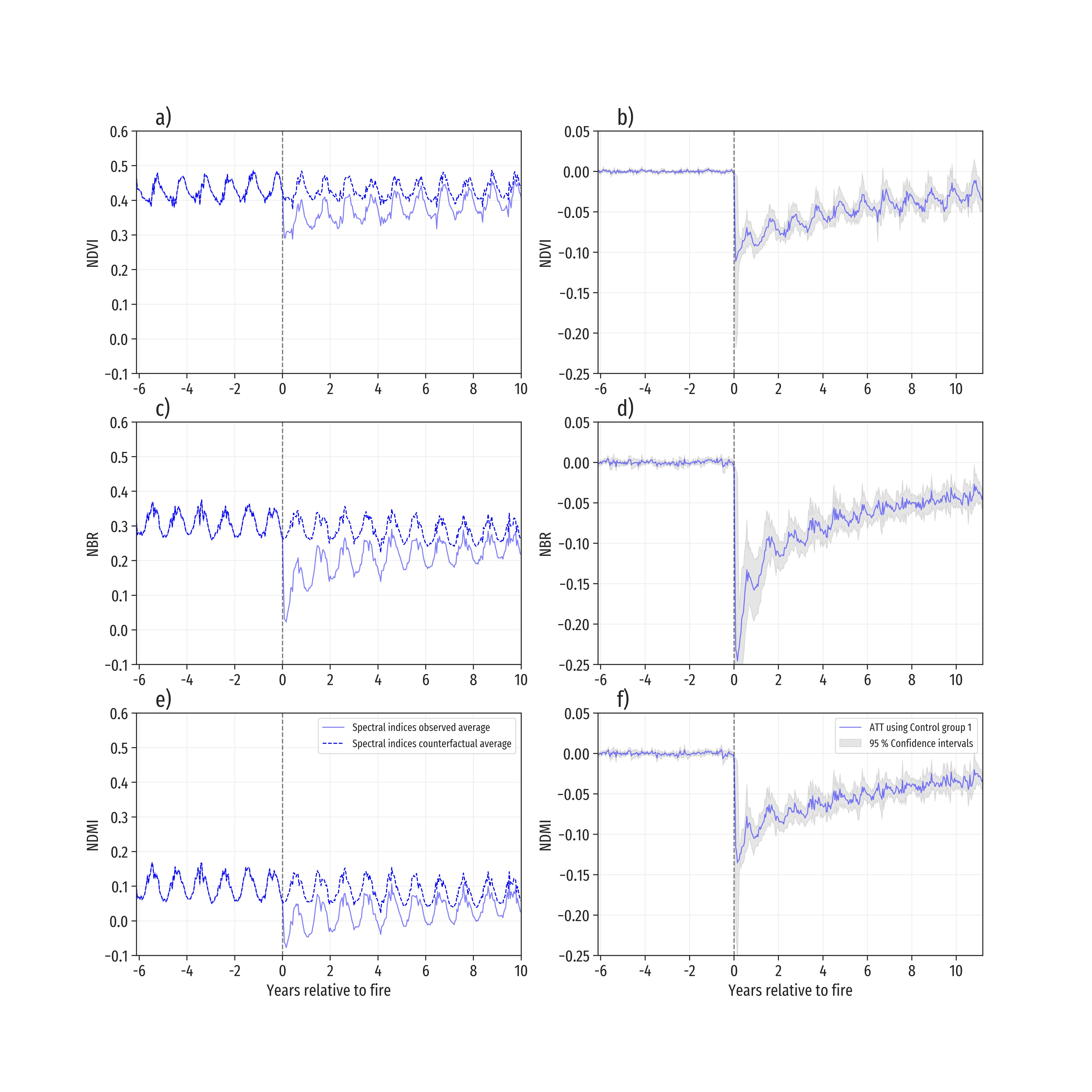}
    \caption{\textbf{Average treatment effect of wildfires on Landsat Surface Reflectance Indices (NDVI, NBR and NDMI).} Effect of wildfires on vegetation health and moisture indices computed using Control Areas 1. The X-axis on all plots shows the years relative to the wildfires, having all wildfires centered at $Years=0$.
    Plots on the left show the averages observed (light blue), together with their counterfactual estimated (dashed dark blue). On the right, figures show the effect estimated using the NNMCM estimation on vegetation and moisture indices. The light grey bands show the standard errors of the estimated effect for each time period estimated using 100 bootstrap iterations with the bootstrap technique explained in the methods section.}
    \label{fig:ATTS_different_controls}
\end{figure}

\begin{figure}[h]
    \includegraphics[scale=.45]{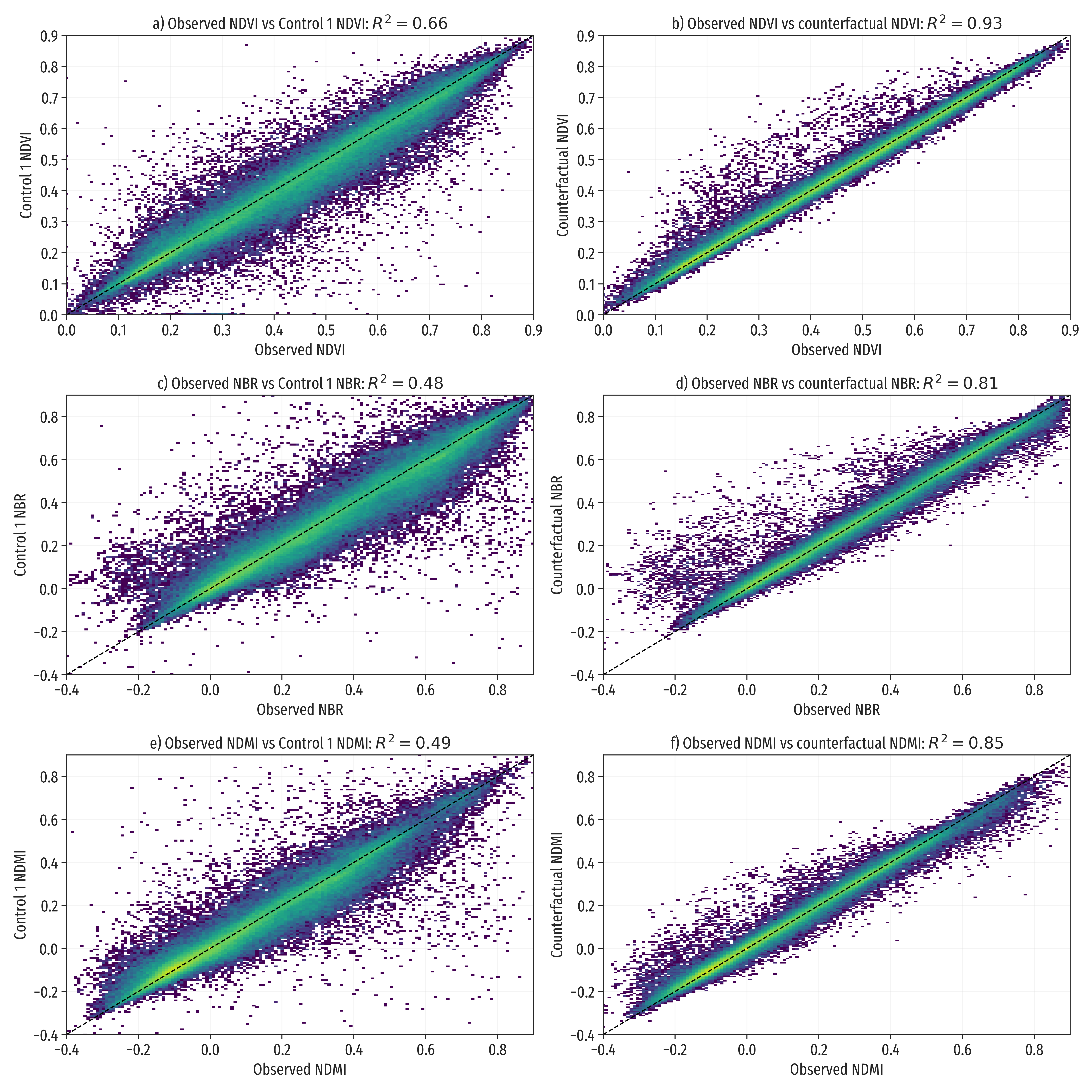}
    \caption{\textbf{Error comparison on pre-wildfire periods of controls and synthetic controls.} 2D Histrogam comparison of the 307 wildfires’ fit on pre-wildfire periods (with minimum $5$ years of pre-wildfire vegetation observed, $\tau=5$), between observed values and control values ((a), (c) and (e)), , and estimated synthetic controls and observed values ((b), (d) and (f)) with a logarithmic color scale. There is an improvement of $R^2$ going from $0.66$ from Control 1 regions to $0.93$ on counterfactual estimations for pre-wildfire periods for NDVI, from $0.48$ to $0.81$ for NBR, and from $0.49$ to $0.85$ for NDMI. However, the estimated counterfactual vegetation seem to over-estimate actual vegetation for some observations.
}
    \label{fig:errors_pre10}
\end{figure}

Observing vegetation values from 1990-2019, the average effect of wildfires from 1996-2016 on NDVI was 
an average initial drop of 25\% (Fig.~\ref{fig:ATTS_different_controls}) over the absolute value of NDVI, and a slow recovery, with an average negative effect up to 10 years after the event of wildfires (Fig. \ref{fig:ATTS_different_controls}(a), (b)).
As we are estimating hypothetical scenarios, our measurement of accuracy is to evaluate the fit on pre-wildfire periods, as well as to visually inspect the hypothetical scenarios. Figure \ref{fig:errors_pre10}(a) shows how the NDVI within the control regions correlates with the NDVI inside the burned regions for pre-wildfire periods ($R^2=0.67)$ and is representative of previous methods \citep[e.g.][]{Goetz2006, Alcaraz-Segura2010, Bolton2015}. Figure~\ref{fig:errors_pre10}(b) shows the correlation between synthetic NDVI values (generated from the GSC method) and the NDVI inside burned regions for the pre-wildfire periods ($R^2=0.93$). Comparison of Figs.~\ref{fig:errors_pre10}(a) and (b) demonstrates that the synthetic NDVI estimated from the GSC method is more accurate than the NDVI derived from the control regions considered in the present study. Similar behaviors follow for NBR and NDMI, going from $0.48$ to $0.81$ and from $0.49$ to $0.85$ respectively.

The Mean Squared Prediction Error (MSPE) for the pre-treatment periods between the counterfactuals estimated and the burned regions for NDVI is $5$ times smaller than the Control region 1 and the burned regions on average. Out of the 307 wildfires studied, only 4 of the areas observed show a larger error of the counterfactual estimated compared to the control regions on pre-wildfire periods. 
For NBR (Fig. \ref{fig:ATTS_different_controls} (c), (d)), and NDMI (Fig. \ref{fig:ATTS_different_controls} (e), (f)), the effect is similar, with NBR showing a larger absolute drop over time. MSPE on pre-treatment periods is more than 5 times lower than using control regions. Furthemore, in all three ATTs estimated, the $\delta_{it}$ on pre-wildfire periods $t < T_{0}^i$ is approximately $0$. The cross validation of $\lambda$ for the NNMCM was done using the \textit{gsynth} package in R, \citep{Rgsynth}.

\subsection{Stratified effect of wildfires}
\begin{figure}[h]
    \includegraphics[scale=.25]{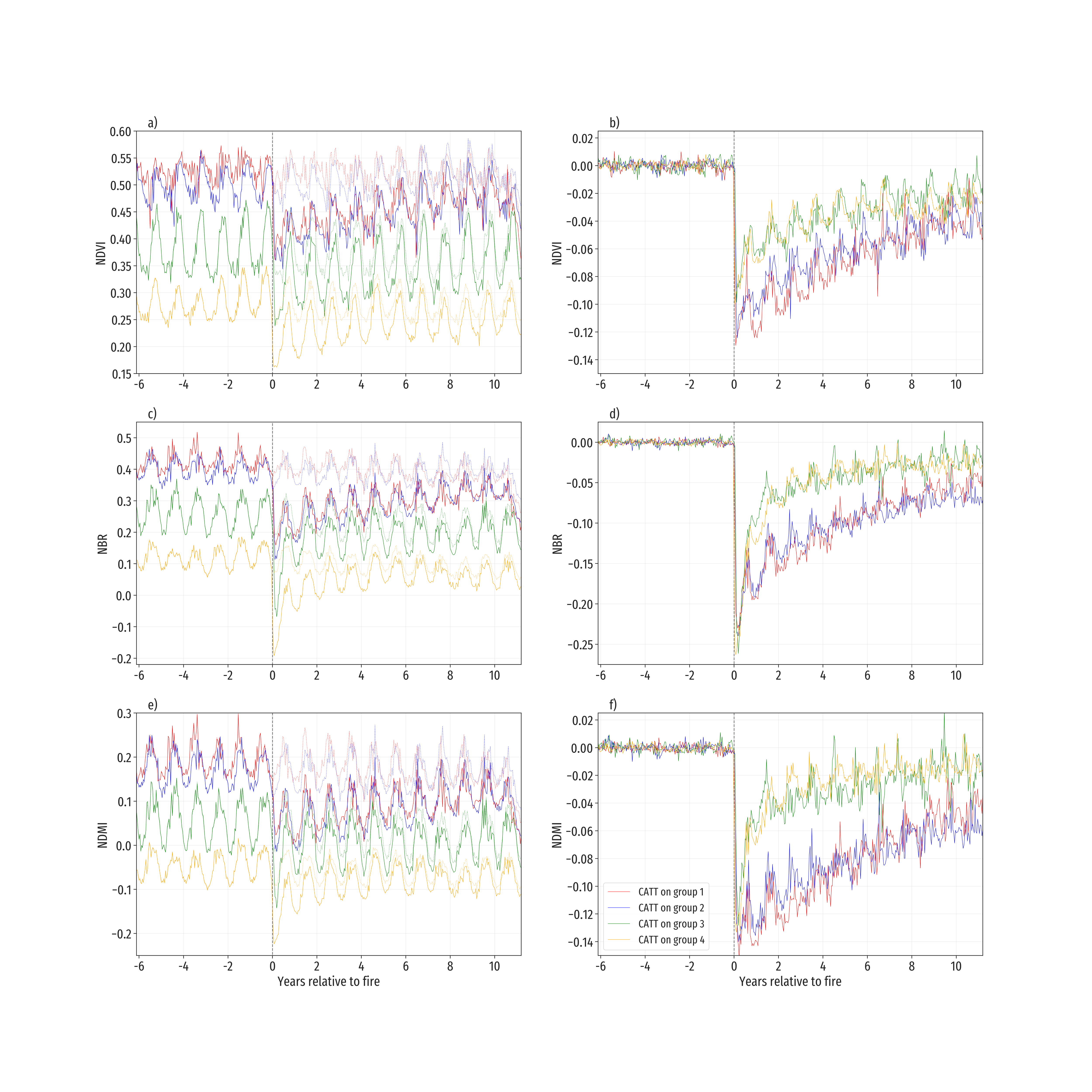}
    \caption{\textbf{Average treatment effect of wildfires stratified based on the BI (i.e. CATT, see Formula (\ref{formula:CATT})).} Conditional Average Treatment Effect on Treated of wildfires on different vegetation and moisture indices, stratifying observations with the burning index from GridMET \citep{Abatzoglou2013}. (a), (c), and (e) show the average NDVI, NBR and NDMI of burned regions respectively (continuous lines), together with their average fitted counterfactuals (dashed lines). We observe that on pre-wildfire periods the fit is accurate, and that the hypothetical counterfactual estimation is insightful. (b), (d), and (f) show the average effect from wildfires for each of the different groups of wildfires. Especifically, AOIs within the groups 3 and 4 of the BI, that happen to have relatively smaller values of NDVI, NBR and NDMI suffer a smaller drop in nominal terms, and the recovery is faster. 
}
    \label{fig:ATTS_stratified_groups}
\end{figure}

The results of the synthetic controls reveal the heterogeneous effects of wildfires in regions with different BI values. Figure ~\ref{fig:ATTS_stratified_groups} shows the average treatment effect of wildfires stratified by BI (groups 1--4 correspond to quartiles 1--4 of the BI). The NDVI, NBR and NDMI averages are markedly different for the four groups for pre-wildfire periods and show different dynamical effects and different recovery patterns post-wildfire. All four groups show a drop in average spectral index value following a wildfire event (Fig.~\ref{fig:ATTS_stratified_groups} (a), (c), (e)). The conditional average effect ($\mathbb{E}(\delta_{it}|X)$) shows that the groups from the lower two quartiles (groups 1 and 2) of the BI (red and blue lines of Fig.~\ref{fig:ATTS_stratified_groups} (b), (d), (f)) suffer a larger change than the upper two quartiles (groups 3 and 4) of the BI (green and yellow lines Fig.~\ref{fig:ATTS_stratified_groups}(b), (d), (f)). 
Furthermore, in all four groups, the average effect on pre-wildfire periods is very close to $0$ (see Fig.~\ref{fig:ATTS_stratified_groups} (b), (d) and (f)), similar to Fig.~\ref{fig:ATTS_different_controls}.
Figures \ref{fig:ATTS_different_controls} and \ref{fig:ATTS_stratified_groups} show that $\hat{\delta}_{it}$ is very close to $0$ for all $t<T_0$ -- that is, we are fulfilling the parallel trends assumption. 

Table \ref{tab:burning_index_globcover} shows that these groups that we used to stratify have different predominant types of vegetation. The Pearson's Chi-squared test gives a p-value of $0.02003$, and thus we can reject the null hypothesis of all groups being equal. Group 1 of the BI is predominated by shrublands and scrublands, Group 2 is predominated by evergreen forests followed closely by shrublands and scrublands, Group 3 is predominated by shrublands and Grasslands, and lastly, Group 4 is predominated by shrublands and grasslands herbaceous observations. Hence, we can discern different recovery patterns from wildfires.

\subsection{Cumulative effect of wildfires}

\begin{figure}[h]
    \includegraphics[scale=.4]{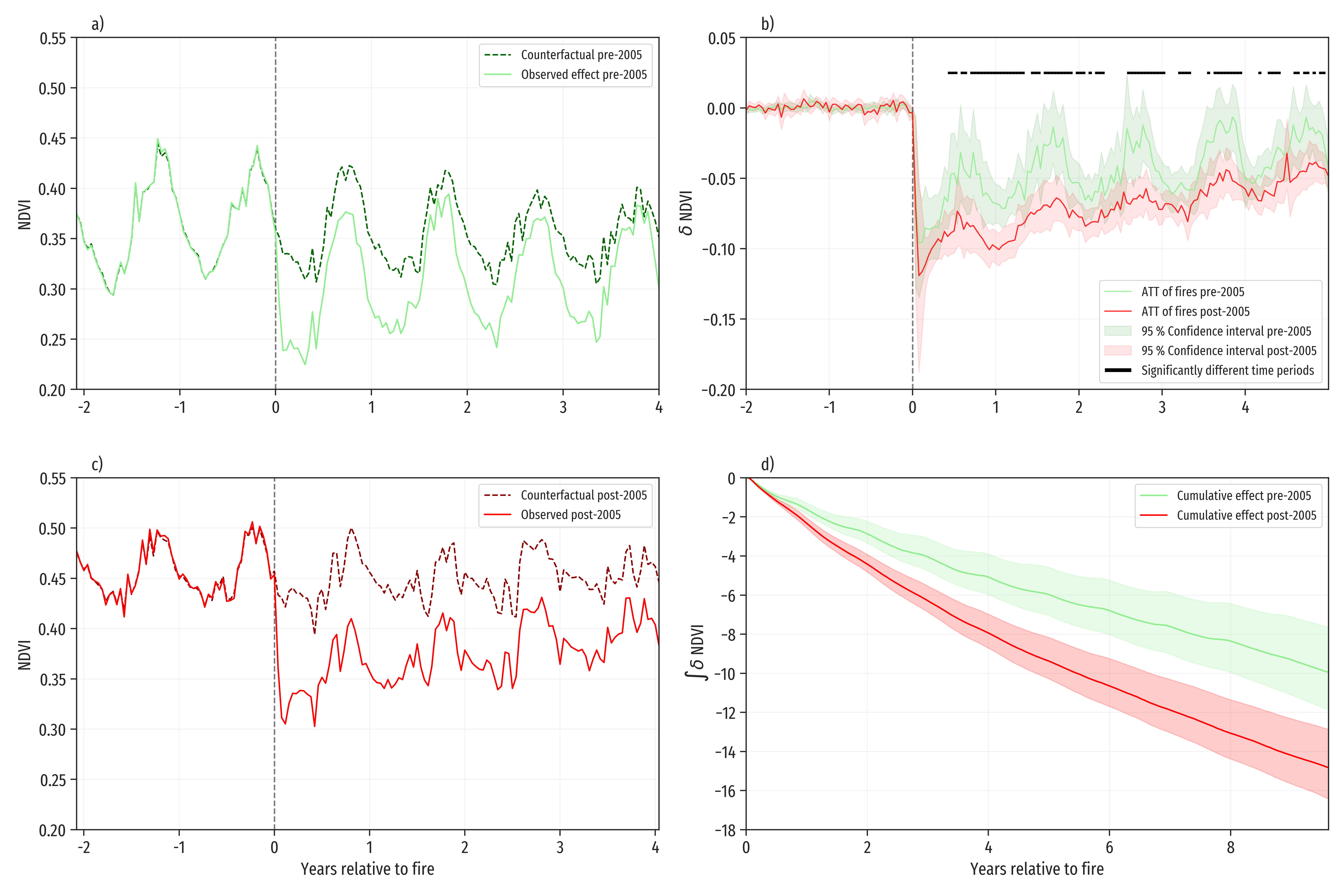}
    \caption{\textbf{Wildfires effect comparison from two decades.} (a) and (c): Average observed and estimated counterfactual NDVI from wildfires from the period 1996-2005 (a), vs average observed and estimated counterfactual NDVI from wildfires on period 2005-2016 (c). (b): Comparison of $ATT_t$ between both time periods, with the respective confidence intervals computed using 100 bootstrap iterations. The dark dashed line above shows the time intervals where the two $ATT_t$ are significantly different, that is, those periods for which 0 does not belong to the $95\%$ normal-based confidence interval of the difference between $ATT_t$s. (d): Comparison of cumulative effects between both periods. Overall, the first decade studied shows a stronger yearly cyclical pattern whereas the second one has larger nominal NDVI values, but less seasonality. The accumulated vegetation loss comparison of the two decades, shows a larger loss for the second period (2005-2016, (d)) than for the first period (1996-2005, (b)). This figure shows changes in the vegetations affected, and the overall change in NDVI over the time period studied, shown in figure \ref{fig:AVG_NDVI}.
}
    \label{fig:cumulative_ATTS}
\end{figure}

Given these long-lasting effects and considering the time period that we are observing, our analysis enables us to not only estimate pre- and post-wildfire differences in spectral indices, but also quantify long-term dynamical changes and compare these. To identify how the effects of wildfires have changed over time, it is instructive to compare subgroups of wildfires across different time-spans \citep{Stevens-Rumann2018}. Specifically, if we divide the whole time-series into two segments (pre- and post-2005), we are able to compare the average treatment effects (Fig.~\ref{fig:cumulative_ATTS} a, b, and c) and average cumulative effects $\sum_{it}\delta_{it}$ (Fig.~\ref{fig:cumulative_ATTS} d) for up to 10 years post-wildfire for both time periods.

From this analysis, we observe three main differences. First, a larger drop in vegetation is detected in the time span 2006-2016 than in 1996-2005 wildfires. Second, a larger difference between the observed average NDVI and the hypothetical vegetation during recovery is also apparent. Last, a less seasonal-fluctuating vegetation, with a more continuous recovery in vegetation is distinguishable. Figure \ref{fig:cumulative_ATTS} (d) also shows an increase in the cumulative effect of wildfires ($\sum_{it}\delta_{it}, t > T_{0}$).

To assess if the changes in seasonal cycles pre- and post-2005 is due to changes in vegetation affected, we have assessed the types of vegetation pre- and post-2005. According to GlobCover 2009 (which has a spatial resolution of 300 m per pixel; See \href{http://due.esrin.esa.int/page_globcover.php}{GlobCover 2009}), it is true that the predominant vegetation types within the burned regions are different between the two time periods, as it can be seen in Table \ref{tab:prepost2005landcover}.

However, these changes do not seem to completely explain this change in terms of the effects of wildfires. If we perform a Pearson’s chi-square test of independence, the p-value is 0.264 and thus we cannot reject the null hypothesis of equal distribution of vegetation pre- and post-2005. One potential explanation of this increase in the effects is the increase in size and quantity of fires that belong to each period. Although both periods have similar size averages, the post-2005 period hast more outliers in terms of size, and the second group contains more non-overlapping wildfires.

\section{Discussion}\label{sec:discussion}


Our results show similar estimates of time recovery from wildfires (from less than 5 years to more than 13 years for recovery, depending on factors as the region, vegetation or environment, among others \citep{Hicke2003, Greene2004, Engel2011}).
Although each burned region represents a distinct effect from the abrupt change caused by wildfires, the GSC method can effectively quantify the loss from wildfires in terms of vegetation health and moisture indices. We have found that the range of effects varies strongly depending on the BI (e.g. Fig.~\ref{fig:ATTS_stratified_groups}), which reflects differences in factors related to fire weather and the ability to control a fire such as topography, fuel moisture, vegetation type and atmospheric conditions. The methodology presented here also allows for precise estimates of the effects of wildfires on vegetation through the estimation of counterfactual hypothetical scenarios in the absence of wildfires. 

Overall, even though a relatively large amount of pre-wildfire periods are required ($\sim 5$ years) to consider seasonalities and trend changes, we find that this methodology could be extended to other fire-prone regions around the world, and that this methodology could be potentially used to estimate the effects of other abrupt changes, always considering the need for control or unaffected regions. 
The assumptions made to construct the models presented in this study are commonly used in statistical procedures \citep{Xu2017}, and this methodology is generalizable as the size of our sample and time period studied are sufficient to detect subtle system changes. 
However, these assumptions could be further validated, for example randomizing treatment with the current data that we have and ensuring that $\delta_{it}$ is flat in pre- and post-wildfire periods. 
For control regions where vegetation is clearly different than the burned regions, counterfactual estimation can be very useful. Hypothetical vegetation for strongly autocorrelated time series with strong seasonalities appear to be harder to model in the long-term. How best to determine when using simpler control regions rather than using synthetic controls is an issue that needs further analysis. 
In this study, models to estimate average treatment effects from wildfires worked best as there were more pre-wildfire periods to calibrate the synthetic controls, and when there were enough observations post-wildfire. Ultimately, the stratification of meaningful groups shows that this methodology allows for identifying different recovery patterns from different types of vegetation and the interaction of this vegetation with different environments. We expect new methodologies showing different stratifications to stem in the future.

\subsection{Evidence for changes in wildfire severity and vegetation recovery}

\begin{figure}[h]
    \includegraphics[scale=.5]{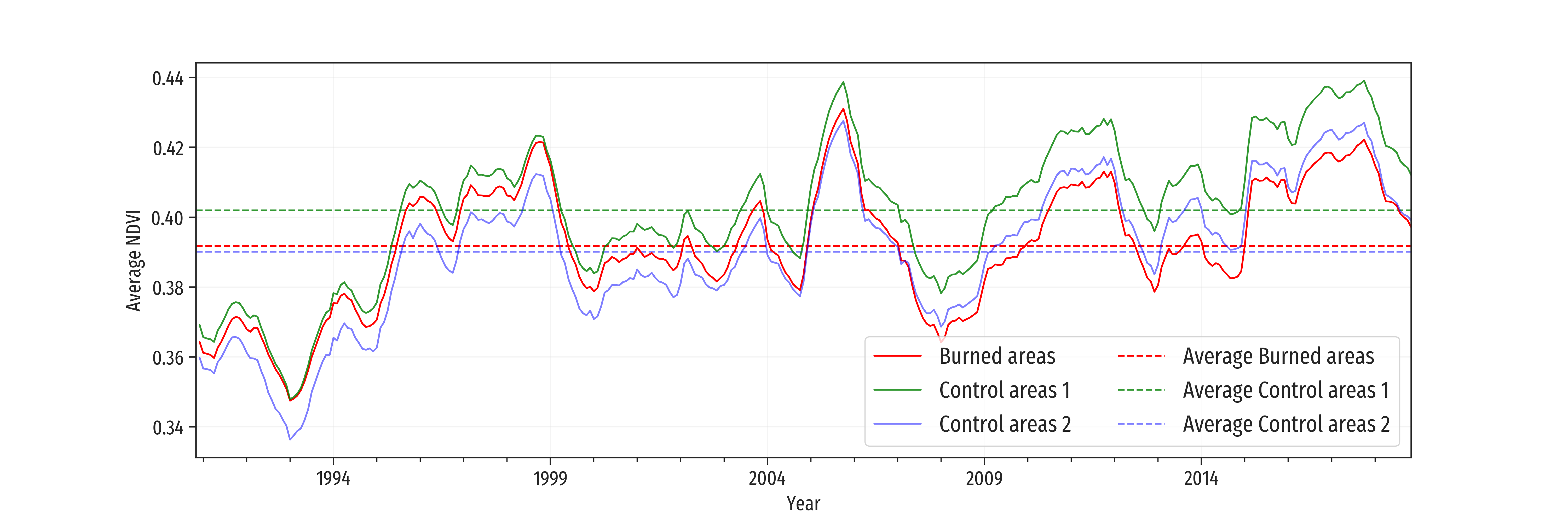}
    \caption{\textbf{Average NDVI over aggregated areas of interest over time.} Average NDVI with a rolling mean of one year. Control region 1 refers to the buffer from 100 meters away from the wildfire to 1 km from it and Control region 2 refers to the regions from 1 km to 5 kms away. Control regions do not include regions overlapping other wildfires, as these were excluded.
}
    \label{fig:AVG_NDVI}
\end{figure}

The composition of ecosystems has undergone constant change over the previous decades  \citep{DALE2001}, and the discrete occurrence of wildfires appears to be increasing \citep{Westerling2006}.
These changes are reflected in our data in several ways.
First, there is a decrease in overall vegetation on the burned areas, compared to non-burned regions (compare red line with both green lines of Control 1 and Control 2 in Fig.~\ref{fig:AVG_NDVI}). Second, we observe that the time series studied are non-stationary, with an upward trend on the average NDVI of the AOIs burned and non-burned, combined with the increase in aridity, potentially resulting in more fuel and conditions for wildfires to burn.
Finally, climate change is affecting fuel flammability in many places, including evergreen forests, and thus, these areas are more prone to burn due to large wildfires \citep{Littell2009}. 

Further research needs to be done to understand how changes in climate are affecting vegetation \citep{Rother2015}. These changes will likely also influence the occurrence of extreme weather events in the future \citep{Westerling2006, Westerling2011, Schoennagel2017, Abatzoglou2016}. The increasing trend on NDVI might be because of sample selection bias, as we are only considering vegetation that burned once, or that did not burn during the time period studied. For many wildfires the long-term effect still remains unclear, but given the size of recent wildfire events, these are expected to be large. 

\subsection{New vegetation cycle patterns}

\begin{figure}[h]
    \includegraphics[scale=.4]{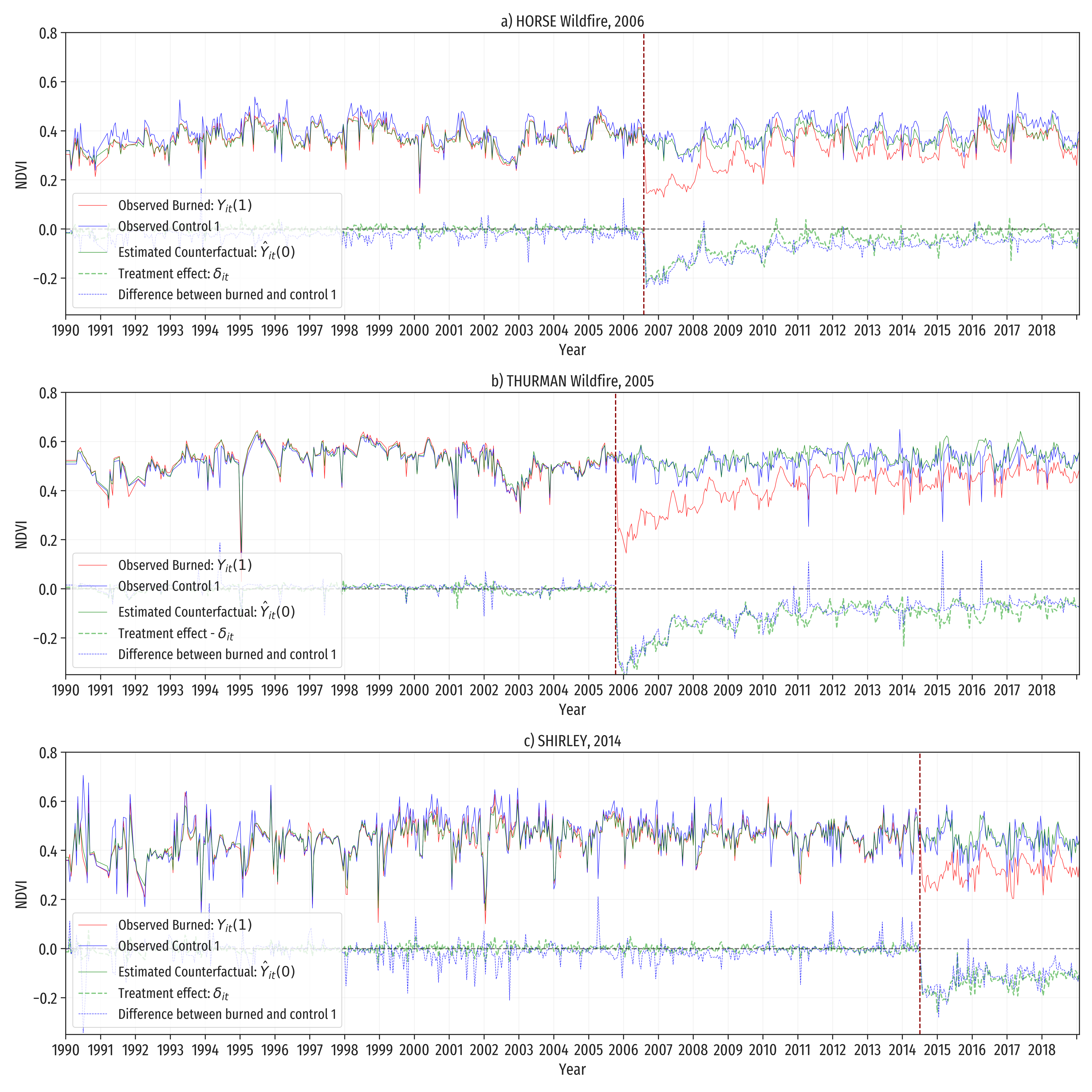}
    \caption{\textbf{Three sample NDVI time series from areas affected by wildfires, together with their respective control region NDVI time series and counterfactual estimates.} 
    (a) Average NDVI over the AOI of wildfire HORSE perimeter, which burned approximately 6766 hectares in 2006. (b) Average NDVI over the AOI of wildfire THURMAN perimeter, which burned approximately 435 hectares in 2005. (c) Average NDVI over the AOI of wildfire SHIRLEY perimeter, which burned approximately 1130 hectares in 2014. Although the three regions had a land cover vegetation predominated by shrubs or scrubs, the effects are different for each wildfire.
}
    \label{fig:Sample_vegetation_aois}
\end{figure}

Figure \ref{fig:Sample_vegetation_aois} shows three examples of how diverse effects of wildfires can be. There are areas with vegetation affected by wildfires having permanent shocks, and never returning to previous states of vegetation, while other areas have recovery periods of less than three years, to more than a decade. This is consistent with previous literature \citep{Hicke2003, Engel2011}, where the impact depends on burn severity, geography, and vegetation. Our analysis enables us to precisely estimate the effects of wildfires.

Wildfires are stochastic events, causing abrupt changes in ecosystems, sometimes resulting in switches to new steady states of vegetation. Figure \ref{fig:Sample_vegetation_aois}(a) shows the Horse Wildfire in 2005, where vegetation recovered after more than 10 years since the wildfire.  The plot also shows how the counterfactual is a better approximation to the observed vegetation than the AOI of Control 1 for this wildfire, as it has been used in previous studies. 
Figure \ref{fig:Sample_vegetation_aois}(b) shows the Thurman wildfire, which seems to not have fully recovered from the wildfire after 14 years. The control region is also appropriate, although the synthetic control shows an even better fit on the pre-wildfire periods.
Lastly, Figure \ref{fig:Sample_vegetation_aois}(c) shows a wildfire where we can not know yet if it will recover because we are only observing 5 years after the wildfire. About a fifth of the wildfires studied cannot be fully evaluated yet, as this highly depends on the phenology and the amount of years after wildfires observed are not enough to determine the amount of years the wildfire will take to recover. Although it could be a potential research line for future work, predictions of whether this vegetation will recover or not are out of the scope of this paper.

\section{Conclusions}\label{sec:conclusions}
The combination of remote sensing data with quasi-experimental observational techniques such as the GSC method has, as far as we know, not been previously used in environmental studies to compute the causal effects of wildfires on vegetation with synthetic controls. Our results demonstrate that the impact of wildfires on vegetation in California is highly variable depending on several factors \citep{Hicke2003, Greene2004, Engel2011}. The estimation of counterfactual vegetation, similar to \citet{Riano2002, Lhermitte2010}, using synthetic controls for larger areas, allows us to observe the effect of each single wildfire, with the help of information collected from the rest of the observations in the same time periods. The nuclear norm matrix completion for synthetic controls method used in this work takes advantage of the information and similarities of observations on pre-wildfire periods, as well as the information from control areas on post-wildfire periods, in order to estimate the effect of wildfires. Hence, the dynamic heterogeneous effects of wildfires can be estimated using Landsat Surface Reflectance - Derived Spectral Indices (LSR-DSI) time-series data.  Future lines of study include expanding and comparing potential areas to analyse, study different vegetation phenologies, estimating new indices \citep{MASSETTI2019167}, and using different satellites or segregation techniques within wildfires. 

Our work illustrates the benefit of using control regions to estimate counterfactual vegetation and assess stochastic shocks in vegetation over time. Even for regions with a decrease in vegetation, we can still infer the hypothetical counterfactual scenario of the absence of a wildfire on a burned region. Considering a proxy of the probability of wildfire occurrence shows that there are discernible patterns from different wildfires.

This methodology allows us to estimate the heterogeneous dynamic causal effects of large wildfires in accordance with climate change \citep{Abatzoglou2016} and seasonal alteration on vegetation health. The main advantages of this approach are:
\begin{itemize}
    \item It is an improvement over previous method as it uses a data-driven methodology for constructing comparable regions to estimate the long-term dynamical wildfire effects.
    \item It allows for different times of wildfire occurrence.
    \item It can be used to estimate counterfactual spectral indices of vegetation and moisture for each of the observed perimeters for a state-wide analysis.
    \item It allows for the estimation of heterogeneity effect estimation of wildfires on spectral indices.
    \item It can be used with more recently developed spectral indices (e.g. VSPI)
    \item It can be applied to a single burned perimeter, using burned pixels and control pixels, and thus detect heterogeneities within a single burned region. 
\end{itemize}
Our results show that, for the three spectral indices considered in the present study (i.e. NDVI, NDMI and NBR), the observed controls surrounding the burned areas are not a good counterfactual (see Figs~\ref{fig:errors_pre10} and \ref{fig:Sample_vegetation_aois}), and that synthetic controls are an effective way of constructing valid counterfactuals (measuring what would have happened in the absence of a fire).
Wildfire disturbance estimation in this study benefits from the inclusion of control regions that are also affected by climate stress. Our results show that places that have lower BI, on average, have larger spectral indices values, such as larger NDVI or larger NDMI values, and also suffer a larger nominal drop after the wildfire, meaning that more vegetation might be lost. 

There are some major challenges and possible improvements to the methods described here. First, as climate is changing, there is an increasing trend on the size of wildfires. Thus, as the frequency of larger events increases, more areas are being burned more than once. As a result, techniques that include multiple treatments or exposures to wildfires could be considered.
Second, better segmentation of different kinds of vegetation would be highly desirable, as these are likely to change differently. As we are only studying large areas being affected by large wildfires, we believe that smaller and targeted areas being burned could also be studied to detect specific types of recoveries for particular types of vegetation.
Third, different types of fires, such as prescribed fires or natural fires allowed to burn when outlined in a fire management plan and communities are not at risk have not been considered here. The methodology that we are using would be able to capture the different effects that these might have on vegetation if sample sizes were large enough.
Fourth, burn severity, which combines the direct fire impact and ecosystems responses \citep{Veraverbeke2010} is heterogeneous within a single wildfire \citep{Bastarrika2011}. Some places are hit harder than others for different reasons (e.g. human land management, vegetation type, etc.). Thus, extending this model to allow for a non-binary or continuous treatment would extend its capacity to reproduce realistic conditions \citep{allofstats2010, Kennedy2017continuous, Powell2020, Hu2020}. In this study we have excluded overlapping wildfires because we wanted to avoid biasing our estimates of counterfactual vegetation. However, studying ways in which we can estimate the effect of a wildfire on vegetation that is recovering from a previous wildfire would further improve this methodology. Finally, the segregation of homogeneous comparable treated and untreated AOIs, that are similar on pre-treatment periods is also expected to follow in future research.

This data-driven methodology is a combination of a quasi-experimental setting with remote sensing data and advanced time series analysis technique for causal inference. Results show that the amount of vegetation lost is increasing, not only because of the increase in the size of wildfires, but also because of the types of vegetation that are being burned. As we want to identify causal effects, we are only considering large wildfire areas that burned once. Moreover, there is an upward trend on NDVI over time over the AOIs. Thus, further studies are required to better understand such abrupt changes in ecosystems as wildfires on vegetation recovery patterns. 

The methodology explained in this study could be applied to other fire-prone regions around the world. 
The computational power of GEE can help further expand this methodology to new regions, allowing to compare wildfire effects over different regions around the world. The only requirement for this would be to find already available perimeters of large wildfires. Replicating this study over many countries could help identify which areas are hit worse by wildfires across the globe, and which kind of vegetation and phenologies are more severely affected by wildfires. Another possible line of future research would be to expand the data used in this study and use this methodology with new spectral indices such as the VSPI \citep{MASSETTI2019167}, or other shocks on vegetation, such as plagues or pests.
Despite the aforementioned challenges, the results point towards significant opportunities ahead. 

\section{Acknowledgements}

We thank all the people that contributed to this research with insightful discussions and comments, including Fred Prata, Miquel Serra-Burriel, Ana Costa-Ramon, Christian Fons-Rosen, Eduardo Graells-Garrido, Patricio Reyes, as well as Guillermo Marin for the help with figures and Victor Paradis for the help editing. Serra-Burriel would also like to thank the Barcelona Supercomputing Center for the Severo Ochoa Mobility Grant, and Delicado would like to thank the Spanish Ministerio de Ciencia e Innovaci\'on for the grant MTM2017-88142-P, and A. T. P. acknowledges funding from the European Union’s Horizon 2020 research and innovation programme under the Marie Skłodowska-Curie grant agreement H2020-MSCA-COFUND-2016-754433. 

The code used in this work has been performed using Python 3.8.1~\citep{10.5555/1593511} and R 3.6.2~\citep{R} programming languages and the Google Earth Engine (GEE) platform \citep{gorelick2017google}. We would also like to acknowledge the following software libraries used in the analysis:  
gsynth~\citep{Rgsynth} (R),
geopandas~\citep{kelsey_jordahl_2020_3946761} (Python), 
numpy~\citep{harris2020array} (Python), 
matplotlib~\citep{hunter2007matplotlib} (Python), 
pandas~\citep{mckinney2011pandas} (Python), 
scipy~\citep{virtanen2020scipy} (Python). 
The code used in this study is available at \url{www.github.com/feliuserra/wildfires\_effects}. 

\section{Description of author's responsibilities}
F. S.-B. led the writing of the paper, produced the figures, downloaded and processed the Landsat data and performed the synthetic controls analysis. P. D. and F. C. contributed to statistical analysis and mathematical formation of the synthetic controls method. A. T. P. contributed to the satellite remote sensing analysis. All authors contributed to the interpretation of results and writing of the manuscript.





\bibliographystyle{elsarticle-num-names}
\bibliography{references.bib}

\begin{thebibliography}{84}
\expandafter\ifx\csname natexlab\endcsname\relax\def\natexlab#1{#1}\fi
\providecommand{\url}[1]{\texttt{#1}}
\providecommand{\href}[2]{#2}
\providecommand{\path}[1]{#1}
\providecommand{\DOIprefix}{doi:}
\providecommand{\ArXivprefix}{arXiv:}
\providecommand{\URLprefix}{URL: }
\providecommand{\Pubmedprefix}{pmid:}
\providecommand{\doi}[1]{\href{http://dx.doi.org/#1}{\path{#1}}}
\providecommand{\Pubmed}[1]{\href{pmid:#1}{\path{#1}}}
\providecommand{\bibinfo}[2]{#2}
\ifx\xfnm\relax \def\xfnm[#1]{\unskip,\space#1}\fi
\bibitem[{Paton et~al.(2015)Paton, Buergelt, Tedim, and McCaffrey}]{paton2015}
\bibinfo{author}{D.~Paton}, \bibinfo{author}{P.~T. Buergelt},
  \bibinfo{author}{F.~Tedim}, \bibinfo{author}{S.~McCaffrey},
\newblock \bibinfo{title}{Wildfires},
\newblock in: \bibinfo{booktitle}{Wildfire {Hazards}, {Risks} and {Disasters}},
  \bibinfo{publisher}{Elsevier}, \bibinfo{year}{2015}, pp.
  \bibinfo{pages}{1--14}. \URLprefix
  \url{https://linkinghub.elsevier.com/retrieve/pii/B9780124104341000014}.
  \DOIprefix\doi{10.1016/B978-0-12-410434-1.00001-4}.
\bibitem[{Easterling et~al.(2000)Easterling, Evans, Groisman, Karl, Kunkel, and
  Ambenje}]{Easterling2000}
\bibinfo{author}{D.~R. Easterling}, \bibinfo{author}{J.~L. Evans},
  \bibinfo{author}{P.~Y. Groisman}, \bibinfo{author}{T.~R. Karl},
  \bibinfo{author}{K.~E. Kunkel}, \bibinfo{author}{P.~Ambenje},
\newblock \bibinfo{title}{{Observed variability and trends in extreme climate
  events: A brief review}},
\newblock \bibinfo{journal}{Bulletin of the American Meteorological Society}
  \bibinfo{volume}{81} (\bibinfo{year}{2000}) \bibinfo{pages}{417--425}.
  \DOIprefix\doi{10.1175/1520-0477(2000)081<0417:OVATIE>2.3.CO;2}.
\bibitem[{Moritz et~al.(2014)Moritz, Batllori, Bradstock, Gill, Handmer,
  Hessburg, Leonard, McCaffrey, Odion, Schoennagel, and Syphard}]{Moritz2014}
\bibinfo{author}{M.~A. Moritz}, \bibinfo{author}{E.~Batllori},
  \bibinfo{author}{R.~A. Bradstock}, \bibinfo{author}{A.~M. Gill},
  \bibinfo{author}{J.~Handmer}, \bibinfo{author}{P.~F. Hessburg},
  \bibinfo{author}{J.~Leonard}, \bibinfo{author}{S.~McCaffrey},
  \bibinfo{author}{D.~C. Odion}, \bibinfo{author}{T.~Schoennagel},
  \bibinfo{author}{A.~D. Syphard},
\newblock \bibinfo{title}{{Learning to coexist with wildfire}},
\newblock \bibinfo{journal}{Nature} \bibinfo{volume}{515}
  (\bibinfo{year}{2014}) \bibinfo{pages}{58--66}.
  \DOIprefix\doi{10.1038/nature13946}.
\bibitem[{Schoennagel et~al.(2017)Schoennagel, Balch, Brenkert-Smith, Dennison,
  Harvey, Krawchuk, Mietkiewicz, Morgan, Moritz, Rasker, Turner, and
  Whitlock}]{Schoennagel2017}
\bibinfo{author}{T.~Schoennagel}, \bibinfo{author}{J.~K. Balch},
  \bibinfo{author}{H.~Brenkert-Smith}, \bibinfo{author}{P.~E. Dennison},
  \bibinfo{author}{B.~J. Harvey}, \bibinfo{author}{M.~A. Krawchuk},
  \bibinfo{author}{N.~Mietkiewicz}, \bibinfo{author}{P.~Morgan},
  \bibinfo{author}{M.~A. Moritz}, \bibinfo{author}{R.~Rasker},
  \bibinfo{author}{M.~G. Turner}, \bibinfo{author}{C.~Whitlock},
\newblock \bibinfo{title}{{Adapt tomore wildfire in western North American
  forests as climate changes}},
\newblock \bibinfo{journal}{Proceedings of the National Academy of Sciences of
  the United States of America} \bibinfo{volume}{114} (\bibinfo{year}{2017})
  \bibinfo{pages}{4582--4590}. \DOIprefix\doi{10.1073/pnas.1617464114}.
\bibitem[{Westerling et~al.(2011)Westerling, Bryant, Preisler, Holmes, Hidalgo,
  Das, and Shrestha}]{Westerling2011}
\bibinfo{author}{A.~L. Westerling}, \bibinfo{author}{B.~P. Bryant},
  \bibinfo{author}{H.~K. Preisler}, \bibinfo{author}{T.~P. Holmes},
  \bibinfo{author}{H.~G. Hidalgo}, \bibinfo{author}{T.~Das},
  \bibinfo{author}{S.~R. Shrestha},
\newblock \bibinfo{title}{{Climate change and growth scenarios for California
  wildfire}},
\newblock \bibinfo{journal}{Climatic Change} \bibinfo{volume}{109}
  (\bibinfo{year}{2011}) \bibinfo{pages}{445--463}.
  \DOIprefix\doi{10.1007/s10584-011-0329-9}.
\bibitem[{Westerling et~al.(2006)Westerling, Hidalgo, Cayan, and
  Swetnam}]{Westerling2006}
\bibinfo{author}{A.~L. Westerling}, \bibinfo{author}{H.~G. Hidalgo},
  \bibinfo{author}{D.~R. Cayan}, \bibinfo{author}{T.~W. Swetnam},
\newblock \bibinfo{title}{{Warming and earlier spring increase Western U.S.
  forest wildfire activity}},
\newblock \bibinfo{journal}{Science} \bibinfo{volume}{313}
  (\bibinfo{year}{2006}) \bibinfo{pages}{940--943}.
  \DOIprefix\doi{10.1126/science.1128834}.
\bibitem[{Spracklen et~al.(2009)Spracklen, Mickley, Logan, Hudman, Yevich,
  Flannigan, and Westerling}]{Spracklen2009}
\bibinfo{author}{D.~V. Spracklen}, \bibinfo{author}{L.~J. Mickley},
  \bibinfo{author}{J.~A. Logan}, \bibinfo{author}{R.~C. Hudman},
  \bibinfo{author}{R.~Yevich}, \bibinfo{author}{M.~D. Flannigan},
  \bibinfo{author}{A.~L. Westerling},
\newblock \bibinfo{title}{{Impacts of climate change from 2000 to 2050 on
  wildfire activity and carbonaceous aerosol concentrations in the western
  United States}},
\newblock \bibinfo{journal}{Journal of Geophysical Research}
  \bibinfo{volume}{114} (\bibinfo{year}{2009}) \bibinfo{pages}{1--17}.
  \DOIprefix\doi{10.1029/2008jd010966}.
\bibitem[{Bryant and Westerling(2014)}]{Bryant2014}
\bibinfo{author}{B.~P. Bryant}, \bibinfo{author}{A.~L. Westerling},
\newblock \bibinfo{title}{{Scenarios for future wildfire risk in California:
  Links between changing demography, land use, climate, and wildfire}},
\newblock \bibinfo{journal}{Environmetrics} \bibinfo{volume}{25}
  (\bibinfo{year}{2014}) \bibinfo{pages}{454--471}.
  \DOIprefix\doi{10.1002/env.2280}.
\bibitem[{Angelo and {Du Plessis}(2017)}]{Angelo2017}
\bibinfo{author}{M.~J. Angelo}, \bibinfo{author}{A.~{Du Plessis}},
\newblock \bibinfo{title}{{Research handbook on climate change and agricultural
  law}},
\newblock \bibinfo{journal}{Research Handbook on Climate Change and
  Agricultural Law}  (\bibinfo{year}{2017}) \bibinfo{pages}{1--472}.
  \DOIprefix\doi{10.4337/9781784710644}.
\bibitem[{Westerling(2016)}]{Westerling2016}
\bibinfo{author}{A.~L.~R. Westerling},
\newblock \bibinfo{title}{{Increasing western US forest wildfire activity:
  Sensitivity to changes in the timing of spring}},
\newblock \bibinfo{journal}{Philosophical Transactions of the Royal Society B:
  Biological Sciences} \bibinfo{volume}{371} (\bibinfo{year}{2016}).
  \DOIprefix\doi{10.1098/rstb.2015.0178}.
\bibitem[{Littell et~al.(2018)Littell, McKenzie, Wan, and
  Cushman}]{Littell2018}
\bibinfo{author}{J.~S. Littell}, \bibinfo{author}{D.~McKenzie},
  \bibinfo{author}{H.~Y. Wan}, \bibinfo{author}{S.~A. Cushman},
\newblock \bibinfo{title}{{Climate Change and Future Wildfire in the Western
  United States: An Ecological Approach to Nonstationarity}},
\newblock \bibinfo{journal}{Earth's Future} \bibinfo{volume}{6}
  (\bibinfo{year}{2018}) \bibinfo{pages}{1097--1111}.
  \DOIprefix\doi{10.1029/2018EF000878}.
\bibitem[{{National Fire Data Center
  (U.S.)}(2005)}]{NationalFireDataCenterU.S.2005}
\bibinfo{author}{{National Fire Data Center (U.S.)}},
\newblock \bibinfo{title}{{The seasonal nature of fires}},
\newblock \bibinfo{journal}{Fema}  (\bibinfo{year}{2005}) \bibinfo{pages}{vi,
  19 p.} \URLprefix
  \url{http://www.usfa.fema.gov/statistics/reports/pubs/seasonal.shtm}.
\bibitem[{Jolly et~al.(2015)Jolly, Cochrane, Freeborn, Holden, Brown,
  Williamson, and Bowman}]{Jolly2015}
\bibinfo{author}{W.~M. Jolly}, \bibinfo{author}{M.~A. Cochrane},
  \bibinfo{author}{P.~H. Freeborn}, \bibinfo{author}{Z.~A. Holden},
  \bibinfo{author}{T.~J. Brown}, \bibinfo{author}{G.~J. Williamson},
  \bibinfo{author}{D.~M. Bowman},
\newblock \bibinfo{title}{{Climate-induced variations in global wildfire danger
  from 1979 to 2013}},
\newblock \bibinfo{journal}{Nature Communications} \bibinfo{volume}{6}
  (\bibinfo{year}{2015}) \bibinfo{pages}{1--11}.
  \DOIprefix\doi{10.1038/ncomms8537}.
\bibitem[{Chu and Guo(2013)}]{Chu2013}
\bibinfo{author}{T.~Chu}, \bibinfo{author}{X.~Guo},
\newblock \bibinfo{title}{{Remote sensing techniques in monitoring post-fire
  effects and patterns of forest recovery in boreal forest regions: A review}},
\newblock \bibinfo{journal}{Remote Sensing} \bibinfo{volume}{6}
  (\bibinfo{year}{2013}) \bibinfo{pages}{470--520}.
  \DOIprefix\doi{10.3390/rs6010470}.
\bibitem[{Hicke et~al.(2003)Hicke, Asner, Kasischke, French, Randerson,
  Collatz, Stocks, Tucker, Los, and Field}]{Hicke2003}
\bibinfo{author}{J.~A. Hicke}, \bibinfo{author}{G.~P. Asner},
  \bibinfo{author}{E.~S. Kasischke}, \bibinfo{author}{N.~H. French},
  \bibinfo{author}{J.~T. Randerson}, \bibinfo{author}{G.~J. Collatz},
  \bibinfo{author}{B.~J. Stocks}, \bibinfo{author}{C.~J. Tucker},
  \bibinfo{author}{S.~O. Los}, \bibinfo{author}{C.~B. Field},
\newblock \bibinfo{title}{{Postfire response of North American boreal forest
  net primary productivity analyzed with satellite observations}},
\newblock \bibinfo{journal}{Global Change Biology} \bibinfo{volume}{9}
  (\bibinfo{year}{2003}) \bibinfo{pages}{1145--1157}.
  \DOIprefix\doi{10.1046/j.1365-2486.2003.00658.x}.
\bibitem[{Meng et~al.(2018)Meng, Wu, Zhao, Cook, Hanavan, and
  Serbin}]{Meng2018}
\bibinfo{author}{R.~Meng}, \bibinfo{author}{J.~Wu}, \bibinfo{author}{F.~Zhao},
  \bibinfo{author}{B.~D. Cook}, \bibinfo{author}{R.~P. Hanavan},
  \bibinfo{author}{S.~P. Serbin},
\newblock \bibinfo{title}{{Measuring short-term post-fire forest recovery
  across a burn severity gradient in a mixed pine-oak forest using multi-sensor
  remote sensing techniques}},
\newblock \bibinfo{journal}{Remote Sensing of Environment}
  \bibinfo{volume}{210} (\bibinfo{year}{2018}) \bibinfo{pages}{282--296}.
  \URLprefix \url{https://doi.org/10.1016/j.rse.2018.03.019}.
  \DOIprefix\doi{10.1016/j.rse.2018.03.019}.
\bibitem[{Harris et~al.(2018)Harris, Beaumont, Vance, Tozer, Remenyi,
  Perkins-Kirkpatrick, Mitchell, Nicotra, McGregor, Andrew, Letnic, Kearney,
  Wernberg, Hutley, Chambers, Fletcher, Keatley, Woodward, Williamson, Duke,
  and Bowman}]{Harris2018}
\bibinfo{author}{R.~M. Harris}, \bibinfo{author}{L.~J. Beaumont},
  \bibinfo{author}{T.~R. Vance}, \bibinfo{author}{C.~R. Tozer},
  \bibinfo{author}{T.~A. Remenyi}, \bibinfo{author}{S.~E. Perkins-Kirkpatrick},
  \bibinfo{author}{P.~J. Mitchell}, \bibinfo{author}{A.~B. Nicotra},
  \bibinfo{author}{S.~McGregor}, \bibinfo{author}{N.~R. Andrew},
  \bibinfo{author}{M.~Letnic}, \bibinfo{author}{M.~R. Kearney},
  \bibinfo{author}{T.~Wernberg}, \bibinfo{author}{L.~B. Hutley},
  \bibinfo{author}{L.~E. Chambers}, \bibinfo{author}{M.~S. Fletcher},
  \bibinfo{author}{M.~R. Keatley}, \bibinfo{author}{C.~A. Woodward},
  \bibinfo{author}{G.~Williamson}, \bibinfo{author}{N.~C. Duke},
  \bibinfo{author}{D.~M. Bowman},
\newblock \bibinfo{title}{{Biological responses to the press and pulse of
  climate trends and extreme events}},
\newblock \bibinfo{journal}{Nature Climate Change} \bibinfo{volume}{8}
  (\bibinfo{year}{2018}) \bibinfo{pages}{579--587}. \URLprefix
  \url{http://dx.doi.org/10.1038/s41558-018-0187-9}.
  \DOIprefix\doi{10.1038/s41558-018-0187-9}.
\bibitem[{Abatzoglou and Williams(2016)}]{Abatzoglou2016}
\bibinfo{author}{J.~T. Abatzoglou}, \bibinfo{author}{A.~P. Williams},
\newblock \bibinfo{title}{{Impact of anthropogenic climate change on wildfire
  across western US forests}},
\newblock \bibinfo{journal}{Proceedings of the National Academy of Sciences of
  the United States of America} \bibinfo{volume}{113} (\bibinfo{year}{2016})
  \bibinfo{pages}{11770--11775}. \DOIprefix\doi{10.1073/pnas.1607171113}.
\bibitem[{Williams et~al.(2019)Williams, Abatzoglou, Gershunov, Guzman-Morales,
  Bishop, Balch, and Lettenmaier}]{Williams2019}
\bibinfo{author}{A.~P. Williams}, \bibinfo{author}{J.~T. Abatzoglou},
  \bibinfo{author}{A.~Gershunov}, \bibinfo{author}{J.~Guzman-Morales},
  \bibinfo{author}{D.~A. Bishop}, \bibinfo{author}{J.~K. Balch},
  \bibinfo{author}{D.~P. Lettenmaier},
\newblock \bibinfo{title}{{Observed Impacts of Anthropogenic Climate Change on
  Wildfire in California}},
\newblock \bibinfo{journal}{Earth's Future} \bibinfo{volume}{7}
  (\bibinfo{year}{2019}) \bibinfo{pages}{892--910}.
  \DOIprefix\doi{10.1029/2019EF001210}.
\bibitem[{Minnich(2018)}]{minnich2018california}
\bibinfo{author}{R.~A. Minnich},
\newblock \bibinfo{title}{California fire climate},
\newblock in: \bibinfo{booktitle}{Fire in California’s Ecosystems},
  \bibinfo{publisher}{University of California Press Oakland, CA},
  \bibinfo{year}{2018}, pp. \bibinfo{pages}{11--25}.
\bibitem[{Goulden et~al.(2012)Goulden, Anderson, Bales, Kelly, Meadows, and
  Winston}]{Goulden2012}
\bibinfo{author}{M.~L. Goulden}, \bibinfo{author}{R.~G. Anderson},
  \bibinfo{author}{R.~C. Bales}, \bibinfo{author}{A.~E. Kelly},
  \bibinfo{author}{M.~Meadows}, \bibinfo{author}{G.~C. Winston},
\newblock \bibinfo{title}{{Evapotranspiration along an elevation gradient in
  California's Sierra Nevada}},
\newblock \bibinfo{journal}{Journal of Geophysical Research: Biogeosciences}
  \bibinfo{volume}{117} (\bibinfo{year}{2012}) \bibinfo{pages}{1--13}.
  \DOIprefix\doi{10.1029/2012JG002027}.
\bibitem[{Casady et~al.(2010)Casady, van Leeuwen, and Marsh}]{Casady2010}
\bibinfo{author}{G.~M. Casady}, \bibinfo{author}{W.~J. van Leeuwen},
  \bibinfo{author}{S.~E. Marsh},
\newblock \bibinfo{title}{{Evaluating Post-wildfire Vegetation Regeneration as
  a Response to Multiple Environmental Determinants}},
\newblock \bibinfo{journal}{Environmental Modeling and Assessment}
  \bibinfo{volume}{15} (\bibinfo{year}{2010}) \bibinfo{pages}{295--307}.
  \DOIprefix\doi{10.1007/s10666-009-9210-x}.
\bibitem[{Calkin et~al.(2020)Calkin, Short, and Traci}]{calkin2020california}
\bibinfo{author}{D.~Calkin}, \bibinfo{author}{K.~Short},
  \bibinfo{author}{M.~Traci},
\newblock \bibinfo{title}{California wildfires},
\newblock \bibinfo{journal}{Emergency management in the 21st century: From
  disaster to catastrophe}  (\bibinfo{year}{2020}) \bibinfo{pages}{155--82}.
\bibitem[{Eidenshink et~al.(2007)Eidenshink, Schwind, Brewer, Zhu, Quayle, and
  Howard}]{Eidenshink2007}
\bibinfo{author}{J.~Eidenshink}, \bibinfo{author}{B.~Schwind},
  \bibinfo{author}{K.~Brewer}, \bibinfo{author}{Z.-L. Zhu},
  \bibinfo{author}{B.~Quayle}, \bibinfo{author}{S.~Howard},
\newblock \bibinfo{title}{{A Project for Monitoring Trends in Burn Severity}},
\newblock \bibinfo{journal}{Fire Ecology} \bibinfo{volume}{3}
  (\bibinfo{year}{2007}) \bibinfo{pages}{3--21}.
  \DOIprefix\doi{10.4996/fireecology.0301003}.
\bibitem[{Li and Banerjee(2020)}]{Li2020}
\bibinfo{author}{S.~Li}, \bibinfo{author}{T.~Banerjee},
\newblock \bibinfo{title}{Spatial and temporal patterns of wildfires in
  california},
\newblock \bibinfo{journal}{Earth and Space Science Open Archive}
  (\bibinfo{year}{2020}) \bibinfo{pages}{20}. \URLprefix
  \url{https://doi.org/10.1002/essoar.10504419.1}.
  \DOIprefix\doi{10.1002/essoar.10504419.1}.
\bibitem[{Zhang et~al.(2018)Zhang, Wang, Li, Cai, Yang, and Yi}]{Zhang2018ndvi}
\bibinfo{author}{Y.~Zhang}, \bibinfo{author}{X.~Wang}, \bibinfo{author}{C.~Li},
  \bibinfo{author}{Y.~Cai}, \bibinfo{author}{Z.~Yang}, \bibinfo{author}{Y.~Yi},
\newblock \bibinfo{title}{{NDVI dynamics under changing meteorological factors
  in a shallow lake in future metropolitan, semiarid area in North China}},
\newblock \bibinfo{journal}{Scientific Reports} \bibinfo{volume}{8}
  (\bibinfo{year}{2018}) \bibinfo{pages}{1--13}. \URLprefix
  \url{http://dx.doi.org/10.1038/s41598-018-33968-w}.
  \DOIprefix\doi{10.1038/s41598-018-33968-w}.
\bibitem[{Dale et~al.(2001)Dale, Joyce, Mcnulty, Neilson, Ayres, Flannigan,
  Hanson, Irland, Lugo, Peterson, Simberloff, Swanson, Stocks, and {Michael
  Wotton}}]{DALE2001}
\bibinfo{author}{V.~H. Dale}, \bibinfo{author}{L.~A. Joyce},
  \bibinfo{author}{S.~Mcnulty}, \bibinfo{author}{R.~P. Neilson},
  \bibinfo{author}{M.~P. Ayres}, \bibinfo{author}{M.~D. Flannigan},
  \bibinfo{author}{P.~J. Hanson}, \bibinfo{author}{L.~C. Irland},
  \bibinfo{author}{A.~E. Lugo}, \bibinfo{author}{C.~J. Peterson},
  \bibinfo{author}{D.~Simberloff}, \bibinfo{author}{F.~J. Swanson},
  \bibinfo{author}{B.~J. Stocks}, \bibinfo{author}{B.~{Michael Wotton}},
\newblock \bibinfo{title}{{Climate Change and Forest Disturbances}},
\newblock \bibinfo{journal}{BioScience} \bibinfo{volume}{51}
  (\bibinfo{year}{2001}) \bibinfo{pages}{723}.
  \DOIprefix\doi{10.1641/0006-3568(2001)051[0723:ccafd]2.0.co;2}.
\bibitem[{Sturrock et~al.(2011)Sturrock, Frankel, Brown, Hennon, Kliejunas,
  Lewis, Worrall, and Woods}]{Sturrock2011}
\bibinfo{author}{R.~N. Sturrock}, \bibinfo{author}{S.~J. Frankel},
  \bibinfo{author}{A.~V. Brown}, \bibinfo{author}{P.~E. Hennon},
  \bibinfo{author}{J.~T. Kliejunas}, \bibinfo{author}{K.~J. Lewis},
  \bibinfo{author}{J.~J. Worrall}, \bibinfo{author}{A.~J. Woods},
\newblock \bibinfo{title}{{Climate change and forest diseases}},
\newblock \bibinfo{journal}{Plant Pathology} \bibinfo{volume}{60}
  (\bibinfo{year}{2011}) \bibinfo{pages}{133--149}.
  \DOIprefix\doi{10.1111/j.1365-3059.2010.02406.x}.
\bibitem[{Seidl et~al.(2017)Seidl, Thom, Kautz, Martin-Benito, Peltoniemi,
  Vacchiano, Wild, Ascoli, Petr, Honkaniemi, Lexer, Trotsiuk, Mairota, Svoboda,
  Fabrika, Nagel, and Reyer}]{Seidl2017}
\bibinfo{author}{R.~Seidl}, \bibinfo{author}{D.~Thom},
  \bibinfo{author}{M.~Kautz}, \bibinfo{author}{D.~Martin-Benito},
  \bibinfo{author}{M.~Peltoniemi}, \bibinfo{author}{G.~Vacchiano},
  \bibinfo{author}{J.~Wild}, \bibinfo{author}{D.~Ascoli},
  \bibinfo{author}{M.~Petr}, \bibinfo{author}{J.~Honkaniemi},
  \bibinfo{author}{M.~J. Lexer}, \bibinfo{author}{V.~Trotsiuk},
  \bibinfo{author}{P.~Mairota}, \bibinfo{author}{M.~Svoboda},
  \bibinfo{author}{M.~Fabrika}, \bibinfo{author}{T.~A. Nagel},
  \bibinfo{author}{C.~P. Reyer},
\newblock \bibinfo{title}{{Forest disturbances under climate change}},
\newblock \bibinfo{journal}{Nature Climate Change} \bibinfo{volume}{7}
  (\bibinfo{year}{2017}) \bibinfo{pages}{395--402}. \URLprefix
  \url{http://dx.doi.org/10.1038/nclimate3303}.
  \DOIprefix\doi{10.1038/nclimate3303}.
\bibitem[{Geist(2005)}]{geist2005our}
\bibinfo{author}{H.~Geist}, \bibinfo{title}{Our earth's changing land: an
  encyclopedia of land-use and land-cover change},
  \bibinfo{publisher}{Greenwood Publishing Group}, \bibinfo{year}{2005}.
\bibitem[{Chuvieco(2012)}]{chuvieco2012remote}
\bibinfo{author}{E.~Chuvieco}, \bibinfo{title}{Remote sensing of large
  wildfires: in the European Mediterranean Basin}, \bibinfo{publisher}{Springer
  Science \& Business Media}, \bibinfo{year}{2012}.
\bibitem[{Goetz et~al.(2006)Goetz, Fiske, and Bunn}]{Goetz2006}
\bibinfo{author}{S.~J. Goetz}, \bibinfo{author}{G.~J. Fiske},
  \bibinfo{author}{A.~G. Bunn},
\newblock \bibinfo{title}{{Using satellite time-series data sets to analyze
  fire disturbance and forest recovery across Canada}},
\newblock \bibinfo{journal}{Remote Sensing of Environment}
  \bibinfo{volume}{101} (\bibinfo{year}{2006}) \bibinfo{pages}{352--365}.
  \DOIprefix\doi{10.1016/j.rse.2006.01.011}.
\bibitem[{Alcaraz-Segura et~al.(2010)Alcaraz-Segura, Chuvieco, Epstein,
  Kasischke, and Trishchenko}]{Alcaraz-Segura2010}
\bibinfo{author}{D.~Alcaraz-Segura}, \bibinfo{author}{E.~Chuvieco},
  \bibinfo{author}{H.~E. Epstein}, \bibinfo{author}{E.~S. Kasischke},
  \bibinfo{author}{A.~Trishchenko},
\newblock \bibinfo{title}{{Debating the greening vs. browning of the North
  American boreal forest: Differences between satellite datasets}},
\newblock \bibinfo{journal}{Global Change Biology} \bibinfo{volume}{16}
  (\bibinfo{year}{2010}) \bibinfo{pages}{760--770}.
  \DOIprefix\doi{10.1111/j.1365-2486.2009.01956.x}.
\bibitem[{Bolton et~al.(2015)Bolton, Coops, and Wulder}]{Bolton2015}
\bibinfo{author}{D.~K. Bolton}, \bibinfo{author}{N.~C. Coops},
  \bibinfo{author}{M.~A. Wulder},
\newblock \bibinfo{title}{{Characterizing residual structure and forest
  recovery following high-severity fire in the western boreal of Canada using
  Landsat time-series and airborne lidar data}},
\newblock \bibinfo{journal}{Remote Sensing of Environment}
  \bibinfo{volume}{163} (\bibinfo{year}{2015}) \bibinfo{pages}{48--60}.
  \URLprefix \url{http://dx.doi.org/10.1016/j.rse.2015.03.004}.
  \DOIprefix\doi{10.1016/j.rse.2015.03.004}.
\bibitem[{Bright et~al.(2019)Bright, Hudak, Kennedy, Braaten, and {Henareh
  Khalyani}}]{Bright2019}
\bibinfo{author}{B.~C. Bright}, \bibinfo{author}{A.~T. Hudak},
  \bibinfo{author}{R.~E. Kennedy}, \bibinfo{author}{J.~D. Braaten},
  \bibinfo{author}{A.~{Henareh Khalyani}},
\newblock \bibinfo{title}{{Examining post-fire vegetation recovery with Landsat
  time series analysis in three western North American forest types}},
\newblock \bibinfo{journal}{Fire Ecology} \bibinfo{volume}{15}
  (\bibinfo{year}{2019}). \DOIprefix\doi{10.1186/s42408-018-0021-9}.
\bibitem[{Ib{\'{a}}{\~{n}}ez et~al.(2019)Ib{\'{a}}{\~{n}}ez, Acharya, Juno,
  Karounos, Lee, McCollum, Schaffer-Morrison, and Tourville}]{Ibanez2019}
\bibinfo{author}{I.~Ib{\'{a}}{\~{n}}ez}, \bibinfo{author}{K.~Acharya},
  \bibinfo{author}{E.~Juno}, \bibinfo{author}{C.~Karounos},
  \bibinfo{author}{B.~R. Lee}, \bibinfo{author}{C.~McCollum},
  \bibinfo{author}{S.~Schaffer-Morrison}, \bibinfo{author}{J.~Tourville},
\newblock \bibinfo{title}{{Forest resilience under global environmental change:
  Do we have the information we need? A systematic review}},
\newblock \bibinfo{journal}{PLoS ONE} \bibinfo{volume}{14}
  (\bibinfo{year}{2019}) \bibinfo{pages}{1--17}.
  \DOIprefix\doi{10.1371/journal.pone.0222207}.
\bibitem[{Kennedy et~al.(2012)Kennedy, Yang, Cohen, Pfaff, Braaten, and
  Nelson}]{Kennedy2012}
\bibinfo{author}{R.~E. Kennedy}, \bibinfo{author}{Z.~Yang},
  \bibinfo{author}{W.~B. Cohen}, \bibinfo{author}{E.~Pfaff},
  \bibinfo{author}{J.~Braaten}, \bibinfo{author}{P.~Nelson},
\newblock \bibinfo{title}{{Spatial and temporal patterns of forest disturbance
  and regrowth within the area of the Northwest Forest Plan}},
\newblock \bibinfo{journal}{Remote Sensing of Environment}
  \bibinfo{volume}{122} (\bibinfo{year}{2012}) \bibinfo{pages}{117--133}.
  \URLprefix \url{http://dx.doi.org/10.1016/j.rse.2011.09.024}.
  \DOIprefix\doi{10.1016/j.rse.2011.09.024}.
\bibitem[{Steiner et~al.(2020)Steiner, Robertson, Teet, Wang, Wu, Zhou, Brown,
  and Xiao}]{Steiner2020}
\bibinfo{author}{J.~L. Steiner}, \bibinfo{author}{S.~Robertson},
  \bibinfo{author}{S.~Teet}, \bibinfo{author}{J.~Wang},
  \bibinfo{author}{X.~Wu}, \bibinfo{author}{Y.~Zhou},
  \bibinfo{author}{D.~Brown}, \bibinfo{author}{X.~Xiao},
\newblock \bibinfo{title}{{Grassland Wildfires in the Southern Great Plains:
  Monitoring Ecological Impacts and Recovery}},
\newblock \bibinfo{journal}{Remote Sensing}  (\bibinfo{year}{2020})
  \bibinfo{pages}{1--15}.
\bibitem[{Lhermitte et~al.(2010)Lhermitte, Verbesselt, Verstraeten, and
  Coppin}]{Lhermitte2010}
\bibinfo{author}{S.~Lhermitte}, \bibinfo{author}{J.~Verbesselt},
  \bibinfo{author}{W.~W. Verstraeten}, \bibinfo{author}{P.~Coppin},
\newblock \bibinfo{title}{A pixel based regeneration index using time series
  similarity and spatial context},
\newblock \bibinfo{journal}{Photogrammetric Engineering and Remote Sensing}
  \bibinfo{volume}{76} (\bibinfo{year}{2010}) \bibinfo{pages}{673--682}.
  \DOIprefix\doi{10.14358/PERS.76.6.673}.
\bibitem[{Xu(2017)}]{Xu2017}
\bibinfo{author}{Y.~Xu},
\newblock \bibinfo{title}{{Generalized synthetic control method: Causal
  inference with interactive fixed effects models}},
\newblock \bibinfo{journal}{Political Analysis} \bibinfo{volume}{25}
  (\bibinfo{year}{2017}) \bibinfo{pages}{57--76}.
  \DOIprefix\doi{10.1017/pan.2016.2}.
\bibitem[{Veraverbeke et~al.(2012)Veraverbeke, Gitas, Katagis, Polychronaki,
  Somers, and Goossens}]{Veraverbeke2012}
\bibinfo{author}{S.~Veraverbeke}, \bibinfo{author}{I.~Gitas},
  \bibinfo{author}{T.~Katagis}, \bibinfo{author}{A.~Polychronaki},
  \bibinfo{author}{B.~Somers}, \bibinfo{author}{R.~Goossens},
\newblock \bibinfo{title}{{Assessing post-fire vegetation recovery using
  red-near infrared vegetation indices: Accounting for background and
  vegetation variability}},
\newblock \bibinfo{journal}{ISPRS Journal of Photogrammetry and Remote Sensing}
  \bibinfo{volume}{68} (\bibinfo{year}{2012}) \bibinfo{pages}{28--39}.
  \URLprefix \url{http://dx.doi.org/10.1016/j.isprsjprs.2011.12.007}.
  \DOIprefix\doi{10.1016/j.isprsjprs.2011.12.007}.
\bibitem[{Veraverbeke et~al.(2010)Veraverbeke, Lhermitte, Verstraeten, and
  Goossens}]{Veraverbeke2010}
\bibinfo{author}{S.~Veraverbeke}, \bibinfo{author}{S.~Lhermitte},
  \bibinfo{author}{W.~W. Verstraeten}, \bibinfo{author}{R.~Goossens},
\newblock \bibinfo{title}{{The temporal dimension of differenced Normalized
  Burn Ratio (dNBR) fire/burn severity studies: The case of the large 2007
  Peloponnese wildfires in Greece}},
\newblock \bibinfo{journal}{Remote Sensing of Environment}
  \bibinfo{volume}{114} (\bibinfo{year}{2010}) \bibinfo{pages}{2548--2563}.
  \URLprefix \url{http://dx.doi.org/10.1016/j.rse.2010.05.029}.
  \DOIprefix\doi{10.1016/j.rse.2010.05.029}.
\bibitem[{Jin et~al.(2019)Jin, Homer, Yang, Danielson, Dewitz, Li, Zhu, Xian,
  and Howard}]{Jin2019}
\bibinfo{author}{S.~Jin}, \bibinfo{author}{C.~Homer},
  \bibinfo{author}{L.~Yang}, \bibinfo{author}{P.~Danielson},
  \bibinfo{author}{J.~Dewitz}, \bibinfo{author}{C.~Li},
  \bibinfo{author}{Z.~Zhu}, \bibinfo{author}{G.~Xian},
  \bibinfo{author}{D.~Howard},
\newblock \bibinfo{title}{{Overall methodology design for the United States
  national land cover database 2016 products}},
\newblock \bibinfo{journal}{Remote Sensing} \bibinfo{volume}{11}
  (\bibinfo{year}{2019}). \DOIprefix\doi{10.3390/rs11242971}.
\bibitem[{Westerling and Wsweinam(2003)}]{Westerling2003}
\bibinfo{author}{A.~L. Westerling}, \bibinfo{author}{T.~Wsweinam},
\newblock \bibinfo{title}{{Interannual to decadal drought and wildfire in the
  Western United States}},
\newblock \bibinfo{journal}{Eos} \bibinfo{volume}{84} (\bibinfo{year}{2003}).
  \DOIprefix\doi{10.1029/2003EO490001}.
\bibitem[{Abatzoglou(2013)}]{Abatzoglou2013}
\bibinfo{author}{J.~T. Abatzoglou},
\newblock \bibinfo{title}{{Development of gridded surface meteorological data
  for ecological applications and modelling}},
\newblock \bibinfo{journal}{International Journal of Climatology}
  \bibinfo{volume}{33} (\bibinfo{year}{2013}) \bibinfo{pages}{121--131}.
  \DOIprefix\doi{10.1002/joc.3413}.
\bibitem[{Hislop et~al.(2018)Hislop, Jones, Soto-Berelov, Skidmore, Haywood,
  and Nguyen}]{Hislop2018}
\bibinfo{author}{S.~Hislop}, \bibinfo{author}{S.~Jones},
  \bibinfo{author}{M.~Soto-Berelov}, \bibinfo{author}{A.~Skidmore},
  \bibinfo{author}{A.~Haywood}, \bibinfo{author}{T.~H. Nguyen},
\newblock \bibinfo{title}{{Using landsat spectral indices in time-series to
  assess wildfire disturbance and recovery}},
\newblock \bibinfo{journal}{Remote Sensing} \bibinfo{volume}{10}
  (\bibinfo{year}{2018}) \bibinfo{pages}{1--17}.
  \DOIprefix\doi{10.3390/rs10030460}.
\bibitem[{Kennedy et~al.(2010)Kennedy, Yang, and Cohen}]{Kennedy2010}
\bibinfo{author}{R.~E. Kennedy}, \bibinfo{author}{Z.~Yang},
  \bibinfo{author}{W.~B. Cohen},
\newblock \bibinfo{title}{{Detecting trends in forest disturbance and recovery
  using yearly Landsat time series: 1. LandTrendr - Temporal segmentation
  algorithms}},
\newblock \bibinfo{journal}{Remote Sensing of Environment}
  \bibinfo{volume}{114} (\bibinfo{year}{2010}) \bibinfo{pages}{2897--2910}.
  \URLprefix \url{http://dx.doi.org/10.1016/j.rse.2010.07.008}.
  \DOIprefix\doi{10.1016/j.rse.2010.07.008}.
\bibitem[{Rouse et~al.(1974)Rouse, Haas, Schell, and
  Deering}]{rouse1974monitoringNDVI}
\bibinfo{author}{J.~Rouse}, \bibinfo{author}{R.~Haas},
  \bibinfo{author}{J.~Schell}, \bibinfo{author}{D.~Deering},
\newblock \bibinfo{title}{Monitoring vegetation systems in the great plains
  with erts},
\newblock \bibinfo{journal}{NASA special publication} \bibinfo{volume}{351}
  (\bibinfo{year}{1974}) \bibinfo{pages}{309}.
\bibitem[{Key and Benson(1999)}]{key1999normalizedBR}
\bibinfo{author}{C.~H. Key}, \bibinfo{author}{N.~C. Benson},
\newblock \bibinfo{title}{The normalized burn ratio (nbr): A landsat tm
  radiometric measure of burn severity},
\newblock \bibinfo{journal}{United States Geological Survey, Northern Rocky
  Mountain Science Center.(Bozeman, MT)}  (\bibinfo{year}{1999}).
\bibitem[{Wilson and Sader(2002)}]{wilson2002detectionNDMI}
\bibinfo{author}{E.~H. Wilson}, \bibinfo{author}{S.~A. Sader},
\newblock \bibinfo{title}{Detection of forest harvest type using multiple dates
  of landsat tm imagery},
\newblock \bibinfo{journal}{Remote Sensing of Environment} \bibinfo{volume}{80}
  (\bibinfo{year}{2002}) \bibinfo{pages}{385--396}.
\bibitem[{Gorelick et~al.(2017)Gorelick, Hancher, Dixon, Ilyushchenko, Thau,
  and Moore}]{gorelick2017google}
\bibinfo{author}{N.~Gorelick}, \bibinfo{author}{M.~Hancher},
  \bibinfo{author}{M.~Dixon}, \bibinfo{author}{S.~Ilyushchenko},
  \bibinfo{author}{D.~Thau}, \bibinfo{author}{R.~Moore},
\newblock \bibinfo{title}{Google earth engine: Planetary-scale geospatial
  analysis for everyone},
\newblock \bibinfo{journal}{Remote sensing of Environment}
  \bibinfo{volume}{202} (\bibinfo{year}{2017}) \bibinfo{pages}{18--27}.
\bibitem[{Cohen et~al.(2017)Cohen, Healey, Yang, Stehman, Brewer, Brooks,
  Gorelick, Huang, Hughes, Kennedy, Loveland, Moisen, Schroeder, Vogelmann,
  Woodcock, Yang, and Zhu}]{Cohen2017}
\bibinfo{author}{W.~B. Cohen}, \bibinfo{author}{S.~P. Healey},
  \bibinfo{author}{Z.~Yang}, \bibinfo{author}{S.~V. Stehman},
  \bibinfo{author}{C.~K. Brewer}, \bibinfo{author}{E.~B. Brooks},
  \bibinfo{author}{N.~Gorelick}, \bibinfo{author}{C.~Huang},
  \bibinfo{author}{M.~J. Hughes}, \bibinfo{author}{R.~E. Kennedy},
  \bibinfo{author}{T.~R. Loveland}, \bibinfo{author}{G.~G. Moisen},
  \bibinfo{author}{T.~A. Schroeder}, \bibinfo{author}{J.~E. Vogelmann},
  \bibinfo{author}{C.~E. Woodcock}, \bibinfo{author}{L.~Yang},
  \bibinfo{author}{Z.~Zhu},
\newblock \bibinfo{title}{{How Similar Are Forest Disturbance Maps Derived from
  Different Landsat Time Series Algorithms?}},
\newblock \bibinfo{journal}{Forests} \bibinfo{volume}{8} (\bibinfo{year}{2017})
  \bibinfo{pages}{1--19}. \DOIprefix\doi{10.3390/f8040098}.
\bibitem[{Roy et~al.(2016)Roy, Kovalskyy, Zhang, Vermote, Yan, Kumar, and
  Egorov}]{Roy2016}
\bibinfo{author}{D.~P. Roy}, \bibinfo{author}{V.~Kovalskyy},
  \bibinfo{author}{H.~K. Zhang}, \bibinfo{author}{E.~F. Vermote},
  \bibinfo{author}{L.~Yan}, \bibinfo{author}{S.~S. Kumar},
  \bibinfo{author}{A.~Egorov},
\newblock \bibinfo{title}{{Characterization of Landsat-7 to Landsat-8
  reflective wavelength and normalized difference vegetation index
  continuity}},
\newblock \bibinfo{journal}{Remote Sensing of Environment}
  \bibinfo{volume}{185} (\bibinfo{year}{2016}) \bibinfo{pages}{57--70}.
  \URLprefix \url{http://dx.doi.org/10.1016/j.rse.2015.12.024}.
  \DOIprefix\doi{10.1016/j.rse.2015.12.024}.
\bibitem[{Zhu and Woodcock(2012)}]{Zhu2012}
\bibinfo{author}{Z.~Zhu}, \bibinfo{author}{C.~E. Woodcock},
\newblock \bibinfo{title}{{Object-based cloud and cloud shadow detection in
  Landsat imagery}},
\newblock \bibinfo{journal}{Remote Sensing of Environment}
  \bibinfo{volume}{118} (\bibinfo{year}{2012}) \bibinfo{pages}{83--94}.
  \URLprefix \url{http://dx.doi.org/10.1016/j.rse.2011.10.028}.
  \DOIprefix\doi{10.1016/j.rse.2011.10.028}.
\bibitem[{Pettorelli et~al.(2005)Pettorelli, Vik, Mysterud, Gaillard, Tucker,
  and Stenseth}]{pettorelli2005usingNDVI}
\bibinfo{author}{N.~Pettorelli}, \bibinfo{author}{J.~O. Vik},
  \bibinfo{author}{A.~Mysterud}, \bibinfo{author}{J.-M. Gaillard},
  \bibinfo{author}{C.~J. Tucker}, \bibinfo{author}{N.~C. Stenseth},
\newblock \bibinfo{title}{Using the satellite-derived ndvi to assess ecological
  responses to environmental change},
\newblock \bibinfo{journal}{Trends in ecology \& evolution}
  \bibinfo{volume}{20} (\bibinfo{year}{2005}) \bibinfo{pages}{503--510}.
\bibitem[{Han et~al.(2011)Han, Ma, Yan, and Song}]{Han2011}
\bibinfo{author}{H.~Han}, \bibinfo{author}{M.~Ma}, \bibinfo{author}{P.~Yan},
  \bibinfo{author}{Y.~Song},
\newblock \bibinfo{title}{{Periodicity analysis of NDVI time series and its
  relationship with climatic factors in the Heihe River Basin in China}},
\newblock \bibinfo{journal}{Remote Sensing for Agriculture, Ecosystems, and
  Hydrology XIII} \bibinfo{volume}{8174} (\bibinfo{year}{2011})
  \bibinfo{pages}{817429}. \DOIprefix\doi{10.1117/12.897938}.
\bibitem[{Running et~al.(1987)Running, Nemani, and Hungerford}]{Running1987}
\bibinfo{author}{S.~W. Running}, \bibinfo{author}{R.~R. Nemani},
  \bibinfo{author}{R.~D. Hungerford},
\newblock \bibinfo{title}{{Extrapolation of synoptic meteorological data in
  mountainous terrain and its use for simulating forest evapotranspiration and
  photosynthesis}},
\newblock \bibinfo{journal}{Canadian Journal of Forest Research}
  \bibinfo{volume}{17} (\bibinfo{year}{1987}) \bibinfo{pages}{472--483}.
  \URLprefix \url{https://doi.org/10.1139/x87-081}.
  \DOIprefix\doi{10.1139/x87-081}.
\bibitem[{National Wildfire Coordinating~Group(2002)}]{Wildfire2002}
\bibinfo{author}{U.~S. D. o.~A. National Wildfire Coordinating~Group},
\newblock \bibinfo{title}{{Gaining an Understanding of the National Fire Danger
  Rating System}}  (\bibinfo{year}{2002}).
\bibitem[{Athey et~al.(2018)Athey, Bayati, Doudchenko, Imbens, and
  Khosravi}]{athey2018matrix}
\bibinfo{author}{S.~Athey}, \bibinfo{author}{M.~Bayati},
  \bibinfo{author}{N.~Doudchenko}, \bibinfo{author}{G.~Imbens},
  \bibinfo{author}{K.~Khosravi}, \bibinfo{title}{Matrix completion methods for
  causal panel data models}, \bibinfo{type}{Technical Report}, National Bureau
  of Economic Research, \bibinfo{year}{2018}.
\bibitem[{{Rubin D. B}(1974)}]{RubinD.B1974}
\bibinfo{author}{{Rubin D. B}},
\newblock \bibinfo{title}{{Estimating causal effects of treatment in randomized
  and nonrandomized studies}},
\newblock \bibinfo{journal}{Journal of Educational Psychology}
  \bibinfo{volume}{66} (\bibinfo{year}{1974}) \bibinfo{pages}{688--701}.
  \URLprefix \url{http://www.fsb.muohio.edu/lij14/420{\_}paper{\_}Rubin74.pdf}.
\bibitem[{Holland(1986)}]{holland1986statistics}
\bibinfo{author}{P.~W. Holland},
\newblock \bibinfo{title}{Statistics and causal inference},
\newblock \bibinfo{journal}{Journal of the American statistical Association}
  \bibinfo{volume}{81} (\bibinfo{year}{1986}) \bibinfo{pages}{945--960}.
\bibitem[{Rosenbaum and Rubin(2006)}]{Rosenbaum2006}
\bibinfo{author}{P.~R. Rosenbaum}, \bibinfo{author}{D.~B. Rubin},
\newblock \bibinfo{title}{{The central role of the propensity score in
  observational studies for causal effects}},
\newblock \bibinfo{journal}{Matched Sampling for Causal Effects}
  (\bibinfo{year}{2006}) \bibinfo{pages}{170--184}.
  \DOIprefix\doi{10.1017/CBO9780511810725.016}.
\bibitem[{Cand{\`{e}}s and Recht(2009)}]{Candes2009}
\bibinfo{author}{E.~J. Cand{\`{e}}s}, \bibinfo{author}{B.~Recht},
\newblock \bibinfo{title}{{Exact matrix completion via convex optimization}},
\newblock \bibinfo{journal}{Foundations of Computational Mathematics}
  \bibinfo{volume}{9} (\bibinfo{year}{2009}) \bibinfo{pages}{717--772}.
  \DOIprefix\doi{10.1007/s10208-009-9045-5}.
\bibitem[{Candes and Plan(2010)}]{candes2010matrix}
\bibinfo{author}{E.~J. Candes}, \bibinfo{author}{Y.~Plan},
\newblock \bibinfo{title}{Matrix completion with noise},
\newblock \bibinfo{journal}{Proceedings of the IEEE} \bibinfo{volume}{98}
  (\bibinfo{year}{2010}) \bibinfo{pages}{925--936}.
\bibitem[{Xu and Liu(2020)}]{Rgsynth}
\bibinfo{author}{Y.~Xu}, \bibinfo{author}{L.~Liu}, \bibinfo{title}{gsynth:
  Generalized Synthetic Control Method}, \bibinfo{year}{2020}. \URLprefix
  \url{http://yiqingxu.org/software/gsynth/gsynth_examples.html},
  \bibinfo{note}{r package version 1.1.7}.
\bibitem[{Stevens-Rumann et~al.(2018)Stevens-Rumann, Kemp, Higuera, Harvey,
  Rother, Donato, Morgan, and Veblen}]{Stevens-Rumann2018}
\bibinfo{author}{C.~S. Stevens-Rumann}, \bibinfo{author}{K.~B. Kemp},
  \bibinfo{author}{P.~E. Higuera}, \bibinfo{author}{B.~J. Harvey},
  \bibinfo{author}{M.~T. Rother}, \bibinfo{author}{D.~C. Donato},
  \bibinfo{author}{P.~Morgan}, \bibinfo{author}{T.~T. Veblen},
\newblock \bibinfo{title}{{Evidence for declining forest resilience to
  wildfires under climate change}},
\newblock \bibinfo{journal}{Ecology Letters} \bibinfo{volume}{21}
  (\bibinfo{year}{2018}) \bibinfo{pages}{243--252}.
  \DOIprefix\doi{10.1111/ele.12889}.
\bibitem[{Greene et~al.(2004)Greene, No{\"{e}}l, Bergeron, Rousseau, and
  Gauthier}]{Greene2004}
\bibinfo{author}{D.~F. Greene}, \bibinfo{author}{J.~No{\"{e}}l},
  \bibinfo{author}{Y.~Bergeron}, \bibinfo{author}{M.~Rousseau},
  \bibinfo{author}{S.~Gauthier},
\newblock \bibinfo{title}{{Recruitment of Picea mariana, Pinus banksiana, and
  Populus tremuloides across a burn severity gradient following wildfire in the
  southern boreal forest of Quebec}},
\newblock \bibinfo{journal}{Canadian Journal of Forest Research}
  \bibinfo{volume}{34} (\bibinfo{year}{2004}) \bibinfo{pages}{1845--1857}.
  \URLprefix \url{https://doi.org/10.1139/x04-059}.
  \DOIprefix\doi{10.1139/x04-059}.
\bibitem[{Engel and Abella(2011)}]{Engel2011}
\bibinfo{author}{E.~C. Engel}, \bibinfo{author}{S.~R. Abella},
\newblock \bibinfo{title}{{Vegetation recovery in a desert landscape after
  wildfires: Influences of community type, time since fire and contingency
  effects}},
\newblock \bibinfo{journal}{Journal of Applied Ecology} \bibinfo{volume}{48}
  (\bibinfo{year}{2011}) \bibinfo{pages}{1401--1410}.
  \DOIprefix\doi{10.1111/j.1365-2664.2011.02057.x}.
\bibitem[{Littell et~al.(2009)Littell, Mckenzie, Peterson, and
  Westerling}]{Littell2009}
\bibinfo{author}{J.~S. Littell}, \bibinfo{author}{D.~Mckenzie},
  \bibinfo{author}{D.~L. Peterson}, \bibinfo{author}{A.~L. Westerling},
\newblock \bibinfo{title}{{Climate and wildfire area burned in western U.S.
  ecoprovinces, 1916-2003}},
\newblock \bibinfo{journal}{Ecological Applications} \bibinfo{volume}{19}
  (\bibinfo{year}{2009}) \bibinfo{pages}{1003--1021}.
  \DOIprefix\doi{10.1890/07-1183.1}.
\bibitem[{Rother et~al.(2015)Rother, Veblen, and Furman}]{Rother2015}
\bibinfo{author}{M.~T. Rother}, \bibinfo{author}{T.~T. Veblen},
  \bibinfo{author}{L.~G. Furman},
\newblock \bibinfo{title}{{A field experiment informs expected patterns of
  conifer regeneration after disturbance under changing climate conditions}},
\newblock \bibinfo{journal}{Canadian Journal of Forest Research}
  \bibinfo{volume}{45} (\bibinfo{year}{2015}) \bibinfo{pages}{1607--1616}.
  \DOIprefix\doi{10.1139/cjfr-2015-0033}.
\bibitem[{Ria{\~{n}}o et~al.(2002)Ria{\~{n}}o, Chuvieco, Ustin, Zomer,
  Dennison, Roberts, and Salas}]{Riano2002}
\bibinfo{author}{D.~Ria{\~{n}}o}, \bibinfo{author}{E.~Chuvieco},
  \bibinfo{author}{S.~Ustin}, \bibinfo{author}{R.~Zomer},
  \bibinfo{author}{P.~Dennison}, \bibinfo{author}{D.~Roberts},
  \bibinfo{author}{J.~Salas},
\newblock \bibinfo{title}{{Assessment of vegetation regeneration after fire
  through multitemporal analysis of AVIRIS images in the Santa Monica
  Mountains}},
\newblock \bibinfo{journal}{Remote Sensing of Environment} \bibinfo{volume}{79}
  (\bibinfo{year}{2002}) \bibinfo{pages}{60--71}.
  \DOIprefix\doi{10.1016/S0034-4257(01)00239-5}.
\bibitem[{Massetti et~al.(2019)Massetti, R{\"{u}}diger, Yebra, and
  Hilton}]{MASSETTI2019167}
\bibinfo{author}{A.~Massetti}, \bibinfo{author}{C.~R{\"{u}}diger},
  \bibinfo{author}{M.~Yebra}, \bibinfo{author}{J.~Hilton},
\newblock \bibinfo{title}{{The Vegetation Structure Perpendicular Index (VSPI):
  A forest condition index for wildfire predictions}},
\newblock \bibinfo{journal}{Remote Sensing of Environment}
  \bibinfo{volume}{224} (\bibinfo{year}{2019}) \bibinfo{pages}{167--181}.
  \URLprefix
  \url{https://www.sciencedirect.com/science/article/pii/S0034425719300586}.
  \DOIprefix\doi{https://doi.org/10.1016/j.rse.2019.02.004}.
\bibitem[{Bastarrika et~al.(2011)Bastarrika, Chuvieco, and
  Mart{\'{i}}n}]{Bastarrika2011}
\bibinfo{author}{A.~Bastarrika}, \bibinfo{author}{E.~Chuvieco},
  \bibinfo{author}{M.~P. Mart{\'{i}}n},
\newblock \bibinfo{title}{{Mapping burned areas from landsat TM/ETM+ data with
  a two-phase algorithm: Balancing omission and commission errors}},
\newblock \bibinfo{journal}{Remote Sensing of Environment}
  \bibinfo{volume}{115} (\bibinfo{year}{2011}) \bibinfo{pages}{1003--1012}.
  \URLprefix \url{http://dx.doi.org/10.1016/j.rse.2010.12.005}.
  \DOIprefix\doi{10.1016/j.rse.2010.12.005}.
\bibitem[{Wasserman(2010)}]{allofstats2010}
\bibinfo{author}{L.~Wasserman}, \bibinfo{title}{All of Statistics: A Concise
  Course in Statistical Inference}, \bibinfo{publisher}{Springer Publishing
  Company, Incorporated}, \bibinfo{year}{2010}.
\bibitem[{Kennedy et~al.(2017)Kennedy, Ma, McHugh, and
  Small}]{Kennedy2017continuous}
\bibinfo{author}{E.~H. Kennedy}, \bibinfo{author}{Z.~Ma},
  \bibinfo{author}{M.~D. McHugh}, \bibinfo{author}{D.~S. Small},
\newblock \bibinfo{title}{{Non-parametric methods for doubly robust estimation
  of continuous treatment effects}},
\newblock \bibinfo{journal}{Journal of the Royal Statistical Society. Series B:
  Statistical Methodology} \bibinfo{volume}{79} (\bibinfo{year}{2017})
  \bibinfo{pages}{1229--1245}. \DOIprefix\doi{10.1111/rssb.12212}.
  \href{http://arxiv.org/abs/1507.00747}{{\tt arXiv:1507.00747}}.
\bibitem[{Powell(2020)}]{Powell2020}
\bibinfo{author}{D.~Powell},
\newblock \bibinfo{title}{{Quantile treatment effects in the presence of
  covariates}},
\newblock \bibinfo{journal}{Review of Economics and Statistics}
  \bibinfo{volume}{102} (\bibinfo{year}{2020}) \bibinfo{pages}{994--1005}.
  \DOIprefix\doi{10.1162/rest_a_00858}.
\bibitem[{Hu et~al.(2020)Hu, Gu, Lopez, Ji, and Wisnivesky}]{Hu2020}
\bibinfo{author}{L.~Hu}, \bibinfo{author}{C.~Gu}, \bibinfo{author}{M.~Lopez},
  \bibinfo{author}{J.~Ji}, \bibinfo{author}{J.~Wisnivesky},
\newblock \bibinfo{title}{{Estimation of causal effects of multiple treatments
  in observational studies with a binary outcome}},
\newblock \bibinfo{journal}{Statistical Methods in Medical Research}
  \bibinfo{volume}{29} (\bibinfo{year}{2020}) \bibinfo{pages}{3218--3234}.
  \DOIprefix\doi{10.1177/0962280220921909}.
\bibitem[{Van~Rossum and Drake(2009)}]{10.5555/1593511}
\bibinfo{author}{G.~Van~Rossum}, \bibinfo{author}{F.~L. Drake},
  \bibinfo{title}{Python 3 Reference Manual}, \bibinfo{publisher}{CreateSpace},
  \bibinfo{address}{Scotts Valley, CA}, \bibinfo{year}{2009}.
\bibitem[{{R Core Team}(2019)}]{R}
\bibinfo{author}{{R Core Team}}, \bibinfo{title}{R: A Language and Environment
  for Statistical Computing}, \bibinfo{organization}{R Foundation for
  Statistical Computing}, \bibinfo{address}{Vienna, Austria},
  \bibinfo{year}{2019}. \URLprefix \url{https://www.R-project.org/}.
\bibitem[{Jordahl et~al.(2020)Jordahl, den Bossche, Fleischmann, Wasserman,
  McBride, Gerard, Tratner, Perry, Badaracco, Farmer, Hjelle, Snow, Cochran,
  Gillies, Culbertson, Bartos, Eubank, maxalbert, Bilogur, Rey, Ren,
  Arribas-Bel, Wasser, Wolf, Journois, Wilson, Greenhall, Holdgraf, Filipe, and
  Leblanc}]{kelsey_jordahl_2020_3946761}
\bibinfo{author}{K.~Jordahl}, \bibinfo{author}{J.~V. den Bossche},
  \bibinfo{author}{M.~Fleischmann}, \bibinfo{author}{J.~Wasserman},
  \bibinfo{author}{J.~McBride}, \bibinfo{author}{J.~Gerard},
  \bibinfo{author}{J.~Tratner}, \bibinfo{author}{M.~Perry},
  \bibinfo{author}{A.~G. Badaracco}, \bibinfo{author}{C.~Farmer},
  \bibinfo{author}{G.~A. Hjelle}, \bibinfo{author}{A.~D. Snow},
  \bibinfo{author}{M.~Cochran}, \bibinfo{author}{S.~Gillies},
  \bibinfo{author}{L.~Culbertson}, \bibinfo{author}{M.~Bartos},
  \bibinfo{author}{N.~Eubank}, \bibinfo{author}{maxalbert},
  \bibinfo{author}{A.~Bilogur}, \bibinfo{author}{S.~Rey},
  \bibinfo{author}{C.~Ren}, \bibinfo{author}{D.~Arribas-Bel},
  \bibinfo{author}{L.~Wasser}, \bibinfo{author}{L.~J. Wolf},
  \bibinfo{author}{M.~Journois}, \bibinfo{author}{J.~Wilson},
  \bibinfo{author}{A.~Greenhall}, \bibinfo{author}{C.~Holdgraf},
  \bibinfo{author}{Filipe}, \bibinfo{author}{F.~Leblanc},
  \bibinfo{title}{geopandas/geopandas: v0.8.1}, \bibinfo{year}{2020}.
  \URLprefix \url{https://doi.org/10.5281/zenodo.3946761}.
  \DOIprefix\doi{10.5281/zenodo.3946761}.
\bibitem[{Harris et~al.(2020)Harris, Millman, van~der Walt, Gommers, Virtanen,
  Cournapeau, Wieser, Taylor, Berg, Smith et~al.}]{harris2020array}
\bibinfo{author}{C.~R. Harris}, \bibinfo{author}{K.~J. Millman},
  \bibinfo{author}{S.~J. van~der Walt}, \bibinfo{author}{R.~Gommers},
  \bibinfo{author}{P.~Virtanen}, \bibinfo{author}{D.~Cournapeau},
  \bibinfo{author}{E.~Wieser}, \bibinfo{author}{J.~Taylor},
  \bibinfo{author}{S.~Berg}, \bibinfo{author}{N.~J. Smith}, et~al.,
\newblock \bibinfo{title}{Array programming with numpy},
\newblock \bibinfo{journal}{Nature} \bibinfo{volume}{585}
  (\bibinfo{year}{2020}) \bibinfo{pages}{357--362}.
\bibitem[{Hunter(2007)}]{hunter2007matplotlib}
\bibinfo{author}{J.~D. Hunter},
\newblock \bibinfo{title}{Matplotlib: A 2d graphics environment},
\newblock \bibinfo{journal}{Computing in Science \& Engineering}
  \bibinfo{volume}{9} (\bibinfo{year}{2007}) \bibinfo{pages}{90}.
\bibitem[{McKinney et~al.(2011)}]{mckinney2011pandas}
\bibinfo{author}{W.~McKinney}, et~al.,
\newblock \bibinfo{title}{pandas: a foundational python library for data
  analysis and statistics},
\newblock \bibinfo{journal}{Python for High Performance and Scientific
  Computing} \bibinfo{volume}{14} (\bibinfo{year}{2011}).
\bibitem[{Virtanen et~al.(2020)Virtanen, Gommers, Oliphant, Haberland, Reddy,
  Cournapeau, Burovski, Peterson, Weckesser, Bright et~al.}]{virtanen2020scipy}
\bibinfo{author}{P.~Virtanen}, \bibinfo{author}{R.~Gommers},
  \bibinfo{author}{T.~E. Oliphant}, \bibinfo{author}{M.~Haberland},
  \bibinfo{author}{T.~Reddy}, \bibinfo{author}{D.~Cournapeau},
  \bibinfo{author}{E.~Burovski}, \bibinfo{author}{P.~Peterson},
  \bibinfo{author}{W.~Weckesser}, \bibinfo{author}{J.~Bright}, et~al.,
\newblock \bibinfo{title}{Scipy 1.0: fundamental algorithms for scientific
  computing in python},
\newblock \bibinfo{journal}{Nature methods} \bibinfo{volume}{17}
  (\bibinfo{year}{2020}) \bibinfo{pages}{261--272}.

\end{thebibliography}

\clearpage

\begin{table}[h]
\begin{tabular}[width=\textwidth]{r|cccc|l}
                                 & \textbf{BI Group 1}   & \textbf{BI Group 2}  & \textbf{BI Group 3}  & \textbf{BI Group 4}   & \textbf{Total} \\ \hline
\textbf{Evergreen Forest}        & 21 (28/24.7\%)    & 28 (37.3/35\%)   & 15 (20/23.8\%)   & 11 (14.7/13.92\%) & 75             \\
\textbf{Grassland Herbaceous}    & 19 (20.2/22.4\%) & 22 (23.4/27.5\%) & 22 (23.4/27.5\%) & 31 (32.98/39.2\%) & 94             \\
\textbf{Shrub/scrublands} & 44 (34.65/51.7\%) & 26 (20.5/32.5\%) & 25 (19.7/39.7\%) & 32 (25.2/40.5\%)  & 127            \\
\textbf{Others}                  & 1 (9/1.2\%)       & 4 (36.4/5\%)     & 1 (9/15.9\%)     & 5 (45.45/6.33\%)  & 11             \\ \hline
\textbf{Total}                   & 85                    & 80                   & 63                   & 79                    & 307           
\end{tabular}
\caption{Table Showing the amount of observations in each category of the Burning Index groups and their predominant landcover according to the GlobCover 2009 dataset. The first number within brackets shows the percentage of the total predominant vegetation and the second number shows the percentage with respect to the total under that BI Group. The Pearson's Chi-squared test gives a p-value of $0.02003$.}
\label{tab:burning_index_globcover}
\end{table}

\begin{table}[h]
    \centering
    \begin{tabular}{r|cc}
         \textbf{Predominant vegetation} & \textbf{Pre-2005} & \textbf{Post-2005} \\ \hline
         \textbf{Evergreen Forest} & 26 (22.03\%) & 49 (25.93\%)\\
         \textbf{Grassland Herbaceous} & 42 (35.6\%) & 52 (27.51\%) \\
         \textbf{Shrub/Scrublands} & 44 (37.3\%) & 83 (43.92\%)\\
         \textbf{Others} & 6 (5.09\%) & 5 (2.65\%)\\ \hline
         \textbf{Total} & 118 & 189
         
    \end{tabular}
    \caption{Comparison of predominant vegetation on pre- and post-2005 perimeters of wildfires. The Pearson's chi-square test of independence between the two periods gives a p-value of 0.264.}
    \label{tab:prepost2005landcover}
\end{table}







\end{document}